\documentstyle[epsfig]{lamuphys}

\newcommand{\refeq}  [1] {(\ref{#1})}

\newcommand{\reffig} [1] {Fig.~\ref{#1}}

\newcommand{\reffigs} [1] {Figs.~\ref{#1}}

\newcommand{\refsect}[1] {Sect.~\ref{#1}}

\newcommand{\refappe}[1] {Appendix~\ref{#1}}

\newcommand{\beq}{\begin{equation}}
\newcommand{\continue}{\nonumber \\ }
\newcommand{\nnu}{\nonumber}
\newcommand{\eeq}{\end{equation}}
\newcommand{\ee}[1] {\label{#1} \end{equation}}
\newcommand{\bea}{\begin{eqnarray}}

\newcommand{\eea}{\end{eqnarray}}
\newcommand{\barr}{\begin{array}}
\newcommand{\earr}{\end{array}}

\newcommand{\MatrixII}[4]{
   \left[
   \begin{array}{cc}
      {#1}  &  {#2}  \\ [1ex]
      {#3}  &  {#4}
   \end{array}
   \right] }

\newcommand{\VectorII}[2]{
   \left[
   \begin{array}{c}
      {#1}  \\ [1ex]
      {#2} 
   \end{array}
   \right] }

\newtheorem{rmark}{{\small\bf\sf Remark}}[chapter]

\newcommand{\Ho}[2]{H^{(1)}_{#1} (k#2)}

\newcommand{\Jb}[2]{J_{#1} (k#2)}
\newcommand{\lesim}{\mbox{\raisebox{-.6ex}{$\,{\stackrel{<}{\sim}}\,$}}}
\newcommand{\gesim}{\mbox{\raisebox{-.6ex}{$\,{\stackrel{>}{\sim}}\,$}}}
\newcommand{\semiclass}{\ \stackrel{\rm s.c.}{\longrightarrow}\ }

\newcommand\threedisk{3-disk repeller}
\newcommand\ndisk{$n$-disk repeller}
\newcommand{\Zqm}{Z_{\rm sc}}			
\newcommand{\Zqc}{Z_{\rm qc}}			
\newcommand{\Vd}{quasi\-classical zeta func\-tion}
\newcommand{\Vt}{quasi\-classical trace formula}

\newcommand{\evOper}{evolution oper\-ator}

\newcommand{\jacobian}{Jacobian}		
\newcommand{\jacobianM}{Jacobian matrix}	
\newcommand{\dzeta}{dyn\-am\-ic\-al zeta func\-tion}

\newcommand{\qS}{semi\-classical zeta func\-tion}
\newcommand{\Gt}{Gutz\-willer trace formula}
\newcommand{\Fd}{Fred\-holm det\-er\-min\-ant}

\newcommand{\ie}{i.e.}

\newcommand{\half}{{\scriptstyle{1\over2}}}
\newcommand{\pde}{\partial}

\renewcommand{\det}{\mbox{\rm det}}
\newcommand{\tr}{{\rm tr}\, }

\newcommand{\prpgtr}[1]{\delta\negthinspace\left( {#1} \right)}

\newcommand{\zfct}[1]{\zeta^{-1}_{#1}}

\newcommand{\Lop}{{\cal L}}
\newcommand{\jMps}{{\bf J}}	   
\newcommand{\jMConfig}{{\bf j}}	   
\newcommand{\jConfig}{j}	   

\newcommand{\maslov}[1]{{m_{#1}}}  

\newcommand\xInit{\xi}			
\newcommand\pSpace{x}		
\newcommand\period[1]{{T_{#1}}}			
\makeatletter
\let\chapter\hid@chapter
\makeatother

\begin{document}
\pagenumbering{arabic}
\title{Quantum Fluids and Classical Determinants}

\author{Predrag\,Cvitanovi\'c\inst{1},  G\'abor\,Vattay\inst{1,2}
	 and Andreas\,Wirzba\inst{3}}

\institute{Center for Chaos and Turbulence Studies,
           Niels Bohr Institute\\ 
	   Blegdamsvej 17, DK-2100 Copenhagen \O, Denmark
\and
E\"otv\"os University, Department of Solid State Physics\\
M\'uzeum krt. 6-8., H-1088 Budapest, Hungary
\and 
Institut f\"ur Kernphysik, TH Darmstadt\\
Schlo{\ss}gartenstra{\ss}e 9, D-64289 Darmstadt, Germany}

\maketitle
\begin{abstract}
A ``quasiclassical'' approximation to the quantum spectrum of
the Schr\"o\-din\-ger equation is obtained from the trace of
a quasiclassical \evOper\ for the ``hydro\-dynamical'' version
of the theory, in which the dynamical evolution
takes place  in the extended phase
space $[q(t),p(t),M(t)] = [q_i, \pde_i S, \pde_i \pde_j S ]$.  
The quasiclassical \evOper\ is multiplicative
along the classical flow, the corresponding \Vd\ is
entire for nice hyperbolic flows, and its eigenvalue spectrum contains the
spectrum of the \qS. The advantage of the \Vd\ is that
it has a larger analyticity domain than the original \qS; the
disadvantage is that it contains eigenvalues extraneous
to the quantum problem. Numerical investigations indicate that
the presence of these extraneous eigenvalues renders 
the original Gutzwiller-Voros \qS\ preferable in 
practice to the \Vd\ presented here. The
cumulant expansion of the exact quantum mechanical scattering kernel 
and the cycle expansion of the corresponding 
semiclassical zeta function part ways at a threshold given by
the topological entropy; beyond this threshold
quantum mechanics cannot resolve fine details of the classical
chaotic dynamics.
\end{abstract}

\section*{Introduction}

What we shall describe here is very much in the spirit of early
quantum mechanics, and were physicists of the period as familiar with
classical chaos as we are today, this theory would have been developed
in 1920's. The main idea is this: in the Bohr--de~Broglie
visualization of quantization, one places a wave instead of a particle
on a Keplerian orbit around the hydrogen nucleus. The quantization
condition is that only allowed orbits are those for which such wave is
stationary; from this follows the Balmer spectrum, the old quantum
theory, and the more sophisticated theory of Schr\"o\-din\-ger and
others. Today we are very aware of the fact that integrable systems
are exceptional and that chaos is the rule. So, can the Bohr
quantization be generalized to chaotic systems?  The answer was
provided by Gutzwiller in 1971; the trace of the quantum \evOper\ for
a chaotic system in a semiclassical approximation is given by the \Gt,
an oscillating sum over all periodic orbits of the system.
 
There is however a hidden intellectual challenge in 
Gutzwiller's derivation: 
the derivation is based on
the semiclassical Van Vleck approximation $K(x,x',t)$ to
the quantum propagator which does not satisfy the semigroup property
\beq
\int \D x''\, K(x,x'',t_1)K(x'',x',t_2)\neq K(x,x',t_1+t_2) \enspace .
\eeq   
In the literature this problem is usually 
sidestepped by saying that an equality holds if the integral is
carried out by the saddle point method. Here we offer an
alternative ``quasiclassical'' quantization scheme based on a
quasiclassical \evOper\ which is multiplicative along
the flow. Our main result is the 
{\em \Vt} for the quantization of a 
Hamiltonian dynamical system. For a system of
2 degrees of freedom the \Vt\ takes the form
\[
\tr {\Lop}^t(E) =
\sum_p  \period{p} \sum_{r=1}^\infty
\frac{
      \prpgtr{t-r\period{p}} 
      \E^{{\I\over \hbar}  (S_p-ET_p) r - \I \pi {\maslov{p} \over 2} r } 
     }{
	    |\Lambda_{p}|^{r/2}
            (1 -{ 1 / \Lambda_{p}^{r} })^2 
            (1 -{ 1 / \Lambda_{p}^{2r} }) 
     }
\enspace .
\]
Throughout this paper we reserve the term ``quasiclassical'' to distinguish
this class of formulae from the original Gutzwiller formulae which 
we shall refer to as ``semiclassical''. 

Search for the above formula was motivated by the classical periodic orbit
theory, where convergence of cycle expansions is under much firmer 
control than in the semiclassical quantizations. One of the main lessons
of the classical theory is that
the ``exponential proliferation of orbits'' in
itself is not the problem; what limits the convergence of cycle
expansions for generic flows is the proliferation of the grammar rules, or the
``algorithmic complexity''.
Indeed, for nice hyperbolic flows a theorem of H.\,H.~Rugh (1992)
asserts that the appropriate spectral determinants are
entire and that their cycle expansions converge superexponentially.

On the basis of close analogy between the classical and the quantum zeta
functions, it has been hoped (Cvitanovi\'c 1992) that for
nice hyperbolic systems the \qS s of \cite{gut84} and \cite{voros88} 
should also be entire.
This hope was dashed by Eckhardt and Russberg (1992) who 
established that the
\qS s for the 3-disk repeller have poles. Their result had in 
turn motivated guesses for spectral determinants with
improved convergence properties by \cite{CR93} and \cite{CRR93},
which eventually lead to the first derivation of the above
trace formula by \cite{CV93}.  In this paper we offer a different
derivation and interpretation of this formula.

Improved analyticity has been very useful in sorting out
the relative importance of the 
semiclassical, diffraction (\cite{Wi92,Wi93}, \cite{VWR94}) 
and quantum contributions
(\cite{GA},  \cite{vattay_BS,vattay_hbar,vattay_ros}).
One had also hoped that improved analiticity would yield
cycle expansions that would converge faster
with the maximal cycle length truncation than the
Gutzwiller-Voros type zeta functions.
As is shown here, this is not the case. Improved analyticity comes
at a cost;  the quasiclassical zeta functions predicts 
extraneous eigenvalues which are purely classical and
do not belong to the quantum spectrum, but 
their presence degrades significantly
the convergence of the cycle expansions.
Furthermore, the investigation of \cite{Wi96} has clarified the 
relationship between the cumulant expansion
of the exact quantum mechanical scattering kernel and the 
cycle expansion of the \qS; the order of expansion at which the two part
their ways is determined by the value of the 
topological entropy, and beyond this threshold quantum mechanics
fails to resolve the arbitrarily fine details of the classical 
chaotic dynamics.

The paper is organized as follows:
in \refsect{s_QMgauge} through \refsect{s_Quasi_class_approx} we develop the 
quasiclassical evolution operator formalism for a
semiclassical approximation to the Schr\"o\-din\-ger equation, and
in \refsect{s_quasi_spectr} we derive the trace and 
zeta function formulae for quasiclassical quantization.
In \refsect{s_num_extran} we confront in numerical experiments
the cycle expansions of the
\Vd s with the cycle expansions of the more standard \qS s
and \dzeta s, as well as with the exact quantum mechanical results,
and in \refsect{s_semi_vs_asymp} we explain the 
distinction between 
the asymptotic nature of quantum mechanical cumulant expansions
and the convergence of semiclassical cycle expansions.
Appendices contain some technical details as well 
as a discussion of the relation of
the quasiclassical quantization to the Selberg zeta function.

\section{Quantum Mechanics in Hydrodynamical Form}  
\label{s_QMgauge}

The Schr\"o\-din\-ger equation for a particle in a $d$-dimensional
potential $V$ is
\beq
\left(
\I \hbar {\partial \over \partial t} 
+ {\hbar^2\over 2} {\Delta}
- V(q)
\right) \psi(q,t) = 0 \enspace ,
\ee{schrod}
where $\psi(q,t)$ is the wave function, and we set the particle
mass $m=1$ throughout.  The ansatz
\beq
\psi = \varphi \E^{\I S/\hbar}  
\ee{ansatz}  
is as old as quantum mechanics itself.
Schr\"o\-din\-ger's first wave mechanics
paper was submitted 27 January 1926.
Submission date for  \cite{Mad1926}
``quantum theory in hydro\-dynamical form'' paper, where this
ansatz is interpreted as a fluid flow,   was 25 October 1926.

Substituting the ansatz into \refeq{schrod}, differentiating, and
separating the result into the real and imaginary parts (under
assumption that both $\varphi$ and $S$ are real functions) yields
\bea
 {\partial S \over \partial t} + {1 \over 2 } 
 \left({\nabla S}\right)^2
 + V(q) - {\hbar^2\over 2}\frac{{\Delta}\varphi}{\varphi} 
 &=& 0 
 \label{Re(Schr)}\\
 {\partial \varphi \over \partial t} 
 + {\nabla S }{\nabla \varphi }
 + {1 \over 2  } {\Delta S }\varphi 
 &=& 0 
 \enspace .
 \label{Im(Schr)}
\eea
The $\hbar^2$ term has many names and is called the ``quantum
potential'' by \cite{Bohm52}, ``enthalpy'' by \cite{speccolo}, by
fluid dynamics analogy, or ``quantum pressure'' by \cite{Feynman_sp}.
While Schr\"o\-din\-ger in his 21 June 1926 paper noted that
$\rho=\varphi\varphi^*$ satisfies the continuity equation, it was Born
who (in a footnote of his 24 june 1926 paper) identified $\rho$ as the
probability density.  Interpretations of quantum mechanics bifurcate
here; keeping the $\hbar$ term in the potential \refeq{Re(Schr)} leads
to the Madelung
``fluid'' theory.
Shifting the $\hbar$ term
into the second  equation enforces that $S$ satisfies the classical 
Hamilton-Jacobi equation
\beq
{\partial S \over \partial t} + {1 \over 2 } 
\left({\nabla S}\right)^2
+ V(q)=0 \enspace , 
\ee{Hamj}
while the ``diffusive'' $\hbar$ term in the equation for the amplitude
\beq
{\partial \varphi \over \partial t} 
+  \nabla S \nabla \varphi 
+ {1\over 2  } {\Delta S }\varphi 
=\frac{\I \hbar}{2}\Delta \varphi
\enspace ,
\ee{trans}
motivates the ``stochastic'' interpretation of \cite{nelson}.

\subsection{Semiclassical Approximation}

Our goal here is to study the semiclassical approximation of quantum
mechanics, with $\hbar$ small, and concentrate on the leading order
expressions. This can be achieved by setting $\hbar$ formally zero in
either the ``hydrodynamic'' or the ``stochastic'' picture.  Either way
we get
\bea
 {\partial S \over \partial t} + {1 \over 2 } 
 \left({\nabla S }\right)^2
 + V(q) &=& 0 
 \label{Hamjj}\\
 {\partial \varphi \over \partial t} 
 +  {\nabla S } 
                     {\nabla \varphi}
 + {1 \over 2  } {\Delta S }\varphi 
 &=& 0
 \enspace .
 \label{cont}
\eea
As long as we concentrate on the leading semiclassical contribution,
we can steer clear of the passions aroused by the differences between
different interpretations of quantum mechanics, and follow the original
Gutzwiller derivation of the semiclassical trace formula via Van-Vleck 
approximation to the quantum propagator, \cite{gut71,gutbook}.
  
Nevertheless, the procedure is unsatisfactory in the sense that in
order to get an operator with the semigroup property we need to impose
the saddle point condition.
In order to overcome this problem we have to learn more about the
technical details of the semiclassical dynamics first. This analysis
will show that the semiclassical wave function evolution can be
described as an evolution over an extended dynamical space.

\section{Semiclassical Evolution as a Set of ODE's}
\label{s_semicl_ev}

We now examine the semiclassical approximation to the quantum wave
evolution (a linear partial differential equation) and show that it
can be reformulated in terms of a finite number of ordinary
differential equations.  We start by traversing a well trodden ground:
Hamilton's 1823 formulation of wave mechanics.

\subsection{Hamilton's Wave Mechanics}
\label{s-Hamilton}

\begin{figure}
\centerline{\epsfig{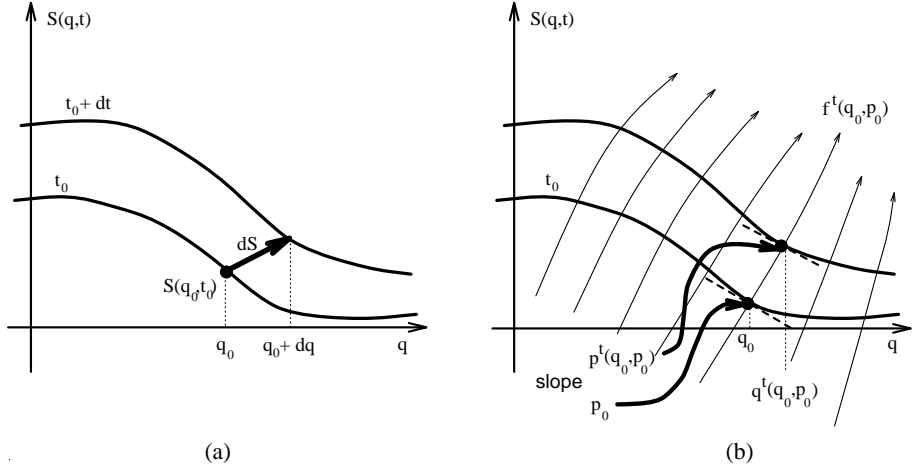}}
\caption[]{
(a) 
A wavefront $S(q,t)$ plotted as a function of
the position $q$ for two different times.
(b) 
The phase of the wavefront $S(q,t)$ transported by a swarm of
``particles''; Hamilton's equations \refeq{Ham_eqs} construct $S(q,t)$
by transporting $q_0 \to q(t)$ and $p_0$, the slope of $S(q_0,t_0)$,
to $p_0 \to p(t)$.  }
\label{f-Ham-char}
\end{figure}

In the wave equation
\refeq{schrod} $q$ is {\em not} a variable; 
the variable is the wave function $\psi$ that evolves with time, and
one can think of $\psi$ as an (infinite dimensional) vector where $q$
plays a role of an index.  $S(q,t)$ plotted as a function of the
position $q$ for two different times looks something like
\reffig{f-Ham-char}(a).
A smooth ``wavefront'' $S(q,t_0)$ deforms smoothly with time into the
``wavefront'' $S(q,t)$ at time $t$. At this point one can ask: could
we think of this front as a swarm of particles that move in such a way
that if we know $S(q,t)$ and its slope $\partial S / \partial q$ at
$q$ at initial time $t=t_0$, we can construct a corresponding piece of
$S(q,t)$ and its slope at time $t$, \reffig{f-Ham-char}(b)?  For
notational convenience, define
\beq
 p_i=p_i(q,t) :={\partial S \over \partial q_i}
 \,, \quad i=1,2,\dots,d
 \enspace .
\ee{(1a)}
In the semiclassical approximation
\refeq{Re(Schr)} reduces to the {\em Hamilton-Jacobi} equation
\beq
 \frac{\partial S}{\partial t} 
  +H\left(q,{\partial S \over \partial q}\right)  = 0
 \enspace ,
\ee{ham_jac}
where $H(q,p)$ is the Hamiltonian, in this case
\beq
  H= p^2/2m + V(q)
\enspace .
\ee{15.6a}
For sake of simplicity we set $m=1$ throughout.
We shall also assume that the Hamiltonian 
is  time independent (energy is conserved) and
separable into a sum of kinetic and potential parts.

Infinitesimal variation of $S(q,t)$, \reffig{f-Ham-char}(a),  
is given by
\[
 \D S = \D t \frac{\partial S}{\partial t} 
     +  \D q {\partial S \over \partial q}
 \enspace .
\]
Dividing through by $\D t$ and substituting \refeq{ham_jac} 
we obtain 
\beq
 {\D S \over \D t}  = - H(q,p) + \dot{q} p
 \enspace .
\ee{dS_dt}
The ``velocity'' $\dot{q}$ is arbitrary, and now comes Hamilton's
idea: can we adjust $\dot{q}$ so that $p$ is promoted to a variable
{\em independent} of $q$? Take a ${\partial~ \over \partial q}$
derivative of both sides of \refeq{dS_dt}:
\[
 {\partial~ \over \partial q} {\D ~ \over \D t} S =
 - {\partial H \over \partial q}
 - {\partial H \over \partial p}{\partial p \over \partial q}
 + p {\partial~ \over \partial q} {\D ~ \over \D t} q + 
  \dot{q} {\partial p \over \partial q}
 \enspace .
\]
(remember that $H(q,p)$ depends on $q$ also through $p(q,t) :=
\partial_q S$, hence the ${\partial H \over \partial p}$ term in the
above).  Exchanging $\partial_q$ and $d/dt$ derivatives leads to
\beq
 \dot{p} +\frac{\partial H}{\partial q} =
 \left(\dot{q} -\frac{\partial H}{\partial p}\right)
 {\partial p \over \partial q}
 \enspace .
\ee{15.7a}
Now we use the freedom of choosing $\dot{q}$, and trade the ${\partial
p \over \partial q}$ dependence for a set of ordinary differential
equations, the Hamilton's equations
\beq
 \dot{q} = \;\;\frac{\partial H(q,p)}{\partial p~~}\enspace , 
 \quad\quad
 \dot{p} = -\frac{\partial H(q,p)}{\partial q~~}
\ee{Ham_eqs}
with the ``wavefront'' $S(q,t)$ replaced by the action 
increment $S^t(q_0,p_0)$,
the integral of \refeq{dS_dt} evaluated along the 
phase space flow $(q_0,p_0) \to (q(t),p(t))$:
\beq
 S^t(q_0,p_0)  = \int_{t_0}^t \D\tau \,
  \left\{ \dot{q}(\tau) \cdot p(\tau) 
                                - H(q(\tau),p(\tau))\right\}
 \enspace .
\ee{S_along_x}
If the energy is conserved, $H(q(\tau),p(\tau))=E$, and the second
term is simply $(t_0-t) E$.

To summarize: 
the Hamilton-Jacobi {\em partial} differential equation
\refeq{ham_jac} for the evolution of a wave front can be reformulated
as a finite number of {\em ordinary} differential equations of motion
which increment the initial action $S(q_0,t_0)$ by the integral
\refeq{S_along_x} along the phase space trajectory
$(q(\tau),p(\tau))$.  In order to obtain the full quasiclassical
evolution we also have to deal with the amplitude evolution
\refeq{cont}.

\subsection{Amplitude Evolution}

The amplitude evolution (\ref{cont}) now takes place in the 
velocity field given by 
\beq
 v(q,t)= \nabla S(q,t) \enspace .
\ee{dina}
We can define $q(t)=f^t(q)$ as a solution of the differential
equation
\beq
 \dot{q}=v(q,t)
\eeq
with initial condition $q(0)=q$ at time $t=0$. This solution will
coincide with $q^t(q,\nabla S(q,0))$, which is the $q$ solution of the
Hamilton's equations with initial conditions $q'=q$ and $p'=\nabla
S(q',0)$.  We introduce the notation $\kappa(q,t)=\Delta S(q,t)$ and
write (\ref{cont}) as
\beq
 \left\{\frac{\partial~}{\partial t}+v(q,t)\cdot\nabla 
 +\frac{1}{2}\kappa(q,t)
      \right\}\varphi(q,t)=0 \enspace .
\ee{cont2}
This is a linear equation in $\varphi$, so its solution can be written
in terms of its Green's function as
\beq
 \varphi(q,t)=\int \D q' \,{\tilde L}^t(q,q')\varphi(q',0)
 \label{phiofq}
\eeq
where the kernel ${\tilde L}^t(q,q')$ is the special solution of
(\ref{cont2}) with initial condition ${\tilde
L}^{0}(q,q')=\delta(q-q')$.  It is easily checked by direct
substitution into
\refeq{phiofq} and \refeq{cont2} 
that
this Green's function is given by
\beq
 {\tilde L}^t(q,q')
  =\exp\left\{\frac{1}{2}\int_0^t\kappa(f^\tau(q'),\tau)
		              \, \D\tau\right\}
 \delta(q-f^{t}(q'))
 \enspace ,
\eeq
where an extra negative contribution to \refeq{cont2}
results  from
$ v(q,t) \nabla \delta(q -f^t(q')) = -(\nabla v(q,t))\delta(q -f^t(q'))$
and $\nabla v(q,t) =\kappa(q,t)$.
 
\subsection{Quasiclassical Evolution}

The whole quasiclassical evolution procedure can now be summarized.
First we take our initial wave function $\psi(q,0)$. We pick a
function $S(q,t)$, a solution of (\ref{Hamjj}), and compute the
initial amplitude $\varphi(q,0)=\E^{-\I S(q,0)/\hbar}\psi(q,0)$. We
evolve this amplitude for time $t$ and put back the phase:
\beq
 \psi(q,t)
  =\E^{\I S(q,t)/\hbar}\int \D q' \,{\tilde L}^t(q,q')
     \E^{-\I S(q',0)/\hbar}\psi(q',0)
 \enspace .
\eeq
The whole evolution can be cast into the semiclassical \evOper
\beq
 \psi(q,t)=\int \D q' \,L^t(q,q',S)\psi(q',0) \enspace ,
\eeq
where
\beq
 L^t(q,q',S)=\exp\left\{\frac{\I}{\hbar}(S(q,t)-S(q',0))
 +\frac{1}{2}\int_0^t\kappa(f^\tau(q'),\tau) \, 
 \D \tau\right\}\delta(q-f^{t}(q'))
 \enspace .
\ee{kern}
The functional dependence on 
$S(q',0)$ sounds
somewhat discouraging; we have to see it in an explicit form in order
to understand the machinery of this operator.

\begin{figure}
\centerline{\epsfig{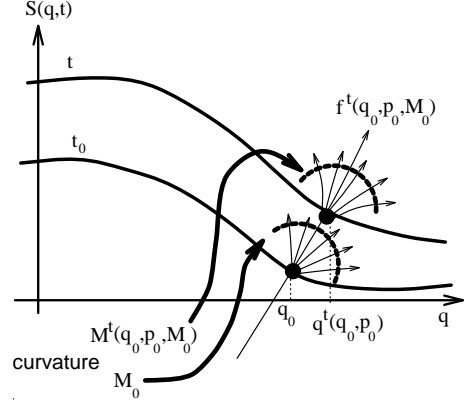}}
\caption[]{
The evolution of the curvature matrix $M_{ij}$ of the wavefront
$S(q,t)$ along the trajectory $[q(t),p(t),M(t)]$ in the extended
dynamical space.
}
\label{f_vattay_curv}
\end{figure}

The most complicated looking object here is the function
\[
 \lambda(q',t)=\int_0^t\kappa(f^\tau(q'),\tau) \, \D\tau
 \enspace .
\] 
We do not need the full information about $S(q,t)$ in order to compute
this integral of $\Delta S(q,t)$ along the trajectory; as we shall
see, an ODE suffices to evaluate this function.  Consider the
curvature matrix
\beq
 {\bf M}_{ij}=\frac{\partial^2 S(q,t)}{\partial q_i\partial q_j}
 \enspace .
\ee{curv_matr}
The time evolution equation for this matrix is obtained by taking the
second derivatives of (\ref{Hamjj}):
\beq
 \frac{\partial {\bf M}}{\partial t}+v(q,t)\cdot\nabla {\bf M} 
 +{\bf M}^2 +{\bf D}^2V =0 \enspace ,
\eeq
where ${\bf D}^2V$ is the second derivative matrix of the potential.
The first two terms combine to the full time derivative, and the
evolution of ${\bf M}$ along a trajectory is given by
\beq
 \dot{{\bf M}}=-{\bf M}^2 -{\bf D}^2V \enspace .
\ee{M_equat}
So in the extended dynamical space we do not only keep track of $q$
and slope of $S$ at $q$, but also the curvature of $S$ at $q$, see
\reffig{f_vattay_curv}.  Let us denote the solution of this ODE along
a trajectory with starting point $(q,p)$ and an initial matrix ${\bf
M}$ by ${\bf M}^t(q,p,{\bf M})$. The function $\lambda(q,t)$ now can
be expressed as
 \beq
 \lambda(q,t)=\int_0^t \D\tau \, \tr {\bf M}^\tau(q,p,{\bf M})  
\eeq
with $p$ initialized as $p=\nabla S(q,0)$.

Another point where the ``functional dependence'' can be simplified is
the phase term. We can make the replacement
\beq
 S(q,t)-S(q',0)=S^t(q',p')
\ee{ide}
in the kernel (\ref{kern}), where $S^t$ is the integral
(\ref{S_along_x}) with initial point $(q',p'=\nabla S(q',0))$ and
$t_0=0$. 

With these observations the kernel (\ref{kern}) can be written as
\bea
 L^t(q,q',S)&=&\int \D p'\, \D {\bf M}'\,\E^{\I S^t(q',p')/\hbar 
 +{1\over 2}\int_0^t \D \tau \,
 \tr {\bf M}^\tau(q',p',{\bf M}')}\times
		\nnu\\
 &&\delta(q-{q}^{t}(q',p'))\,\delta(p'-\nabla S(q',0))
  	\,\delta({\bf M}'-{\bf D}^2 S(q',0))
 \enspace ,
 \label{interm_kern}
\eea
where we have made the functional dependence explicit.

\section{Quasiclassical Evolution Operator}
\label{s_Quasi_class_approx}

If we write the time evolution of a wave function we get
\begin{eqnarray}
 &&\psi(q,t)=\int \D q'\,\D p'\, \D {\bf M}'\,W^t(q',p',{\bf M}')\,
 \delta(q-q^t(q',p')) \nonumber\\
 &&\times\psi(q',0)\,
 \delta(p'-\nabla S(q',0))\,\delta({\bf M}'-{\bf D}^2 S(q',0)) \enspace ,
 \label{fele}
\end{eqnarray}
where $W^t(q,p,{\bf M})$ is a short hand notation for the exponential
in \refeq{interm_kern}.  We now make a new proposal: let us regard the
last deltas $\delta(p'-\nabla S(q',0))\,\delta({\bf M}'-{\bf D}^2
S(q',0))$ as a part of the wave function. In other words, we think
$\Psi(q',p',{\bf M}')=\psi(q',0)\,\delta(p'-\nabla S(q',0))\,
\delta({\bf M}'-{\bf D}^2 S(q',0))$
as a function defined on the $(q,p,{\bf M})$ space.
We can multiply \refeq{fele} by
$\delta(p-\nabla S(q,t)) \,\delta({\bf M}-{\bf D}^2 S(q,t))$
and write the evolved function in the extended space as
\[
 \Psi(q,p,{\bf M})=\int \D q'\,\D p'\,\D {\bf M}'\,
 {\cal L}^t(q,p,{\bf M}|q',p',{\bf M}')\,
 \Psi(q',p',{\bf M}')
 \enspace ,
\]
where the kernel of this integral operator shall be referred to as the
{\em quasiclassical \evOper}
\begin{eqnarray}
 &&{\cal L}^t(q,p,{\bf M}|q',p',{\bf M}')=\E^{\I S^t(q',p')/\hbar 
 +{1\over 2}\int_0^t \D \tau\,
 \tr {\bf M}^\tau(q',p',{\bf M}')}\nonumber \\
 &&\times  \delta(q-q^t(q',p'))\,\delta(p-p^t(q',p'))
    \,\delta({\bf M}-{\bf M}^t(q',p',{\bf M}'))
 \enspace .
 \label{a_main_res}
\end{eqnarray}
Here the quantities $\nabla S(q,t)$ and ${\bf D}^2S(q,t)$ are computed
from their initial values and replaced with $p^t(q',p')$ and 
${\bf M}^t(q',p',{\bf M}')$ using (\ref{ide}).

So, what does this mean? We have constructed an \evOper\ which acts on
functions of the $(q,p,{\bf M})$ space. Because of the three 
delta functions the \evOper\ 
has the semigroup property. 
However, there will be a price to pay: while a wave function can
be embedded into the enlarged space, not all the functions living in
the enlarged space represent functions in the old space.  The spectrum
of the quasiclassical operator will contain the semiclassical
spectrum, but as we shall see in \refsect{s_num_extran}, it will also
contain extraneous eigenvalues without quantum mechanical counterpart.

\subsection{Wave Packet Evolution}

There is also an easy way back from the extended space
to the original one.  If the function $\Psi(q,p,{\bf M})$ is a
representation of a $q$ space wave function or represents a linear
combination of such functions, the delta function dependence on $p$
and ${\bf M}$ insures that a $q$ dependent wave function can be
recovered by
\beq
 \psi(q,t)=\int \D p\, \D {\bf M} \, \Psi^t(q,p,{\bf M})
 \enspace .
\eeq
The quasiclassical evolution introduced here is closely related to the
Gaussian wave packet evolution theories of \cite{Heller75,Heller92}.  
There a
wave packet
\beq
 \psi(q,0)=A_0\E^{\I
 p_0(q-q_0)/\hbar+\frac{\I}{2\hbar}(q-q_0){\bf M}_0(q-q_0)}
\eeq
is ``launched" at $t=0$, with the parameters $(q_0,p_0,{\bf M}_0)$
evolving in time according to the equations we have for $q$, $p$ and
${\bf M}$, and with the amplitude evolving as
\beq
 A^t=A_0\E^{\I S^t(q_0,p_0)/\hbar-\frac{1}{2}\int_0^t \D\tau\, 
 \tr {\bf M}^{\tau}(q_0,p_0,{\bf M}_0)}
\enspace .
\eeq
Initial wave functions can be decomposed into a linear combination of
wave packets and the pieces can be evolved separately. Each packet is
characterized by a phase point in the ($q,p,{\bf M}$) phase space and
evolves according to \refeq{Ham_eqs} and \refeq{M_equat}, with clouds
of points representing initial wave packets evolving as in the
\cite{Heller92} picture.

\subsection{A Classical Motivation for the Extended Dynamical Space}

The above discussion might lead the reader to believe that the
extended dynamical phase space is a peculiarity of quantum
quasiclassics.  However, what we have done is an example of a much
more general procedure for constructing multiplicative \evOper s in
settings where the multiplicative property seems to have been lost.

The problem can be illustrated by the \cite{Ruelle} ``thermodynamic"
\evOper\ of form
\[
 {\cal L}^t(x,x')=
 \E^{h^t (x')}\,
 \delta(x-f^t(x'))\,
 {1 \over |\Lambda^t(x')|^{\beta-1} }
 \enspace ,
\]
with $\Lambda^t(x)$ an eigenvalue of the Jacobi matrix ${\bf J}^t(x)$
(see Appendix A) and $h^t(x)$ is a weight additive along the
trajectory $f^t(x)$.  For one-dimensional maps
this operator is multiplicative, but {\em not} so for flows with two
or more transverse dimensions, for the simple reason that the
eigenvalues of successive stability matrices are in general not
multiplicative
\[
 \Lambda_{ab} \neq  \Lambda_a \Lambda_b \enspace .
\]
Here ${\bf J}_{ab}={\bf J}_{b}{\bf J}_{a}$ is the
\jacobianM\ of the trajectory
consisting of consecutive segments $a$ and $b$, ${\bf J}_{a}$ and
${\bf J}_{b}$ are the stability matrices for these segments
separately, and $\Lambda$'s are their leading eigenvalues. It was this
lack of multiplicative property for $\Lambda$'s that had for long time
frustrated attempts to construct \evOper s whose spectrum contains the
semiclassical Gutzwiller spectrum, until the method presented here was
developed.

The main idea, extending the dynamical system to the tangent space of
the flow, is suggested by one of the standard numerical methods for
evaluation of Lyapunov exponents; instead of computing eigenvalues of
linearized stability matrices, one monitors the growth rate of
separation between nearby trajectories, \ie\ one adjoins
space $x\in U\subset {\bbbr}^{d}$.  The dynamics in the $(x,\xi) \in U
\times TU_x$ space is governed by the system of equations of
variations, \cite{arnold73}:
\[
 \dot{x}={\bf v}(x) \,, \quad  
 \dot{{\bf \xi}}={\bf Dv}(x){\bf \xi }\enspace .
\]
Here ${\bf Dv}(x)$ is the derivative matrix of the flow.
We write the solution as
\[
 x(t)=f^t(x_0) \enspace , \quad 
 {\bf \xi}(t)={\bf J}^t(x_0) \cdot {\bf \xi}_0 
\] 
with the tangent space vector ${\bf \xi}$ transported by the
transverse stability matrix ${\bf J}^t(x_0) = \partial x(t)/ \partial
x_0$.  Multiplicative \evOper s and the corresponding trace and
determinant formulae for such flows are given in \cite{CV93} and 
\cite{Pollner}.

\section{Quasiclassical Trace and Determinant Formulae}
\label{s_quasi_spectr}

Determination of the approximate eigenvalues of the Schr\"o\-din\-ger
operator \refeq{schrod} is now reduced to the computation of the
eigenvalues of the quasiclassical
\evOper\ \refeq{a_main_res}. 
But before we do this, a warning is in order.  The spectrum of the new
operator {\em contains} the semiclassical spectrum, \ie\ we might find
eigenvalues beyond those found in quantum mechanics.  Optimally these
extraneous eigenvalues should be filtered out, but at present we know
of no practical technique for doing this, other than comparison with
the exact quantum mechanical spectrum.
 
We shall determine the eigenvalues of our operator by first deriving
the classical trace formula (\cite{CEflows}, \cite{QCcourse}), and
then determining the zeros of the associated \Fd, in this context
called the \Vd.  The $(p,q)$ integrations can be carried out first,
and yield a weighted sum over primitive periodic orbits $p$ and their
repetitions $r$
\beq
 \tr {\Lop}^t(E) =
 \sum_p  \period{p} \sum_{r=1}^\infty 
 \frac{\prpgtr{t-r\period{p}} \E^{{\I\over \hbar}  (S_p-ET_p) r} }
     {\mid \det(1-{\bf J}_p^r)\mid} 
 \Delta_{p,r} \enspace .
\ee{Tr_La1}
By the periodicity condition $\prpgtr{t-r\period{p}}$ the ${\bf M}$
trace is restricted to a transverse Poincar\'e section of the flow,
evaluated at a prime cycle completion $t=\period{p}$, or its $r$-th
repeat
\beq
 \Delta_{p,r} = \int \D{\bf M} \,
 \prpgtr{{\bf M}-{\bf M}^{r\period{p}}(q,p,M)}
 \E^{\frac{r}{2}\int_0^{T_p}\D\tau \tr{\bf M}^\tau(q,p,{\bf M})} 
 \enspace .
\ee{Tr_La2}
The integration of this part requires some skill and it is left for
\refappe{App_Jacob}.  It turns out that this last integral can also be
expressed in terms of the eigenvalues of the full phase space
\jacobianM\ $\Lambda_{1}, \Lambda_{2}, \cdots,
\Lambda_{d+1}\mbox{=}
1/\Lambda_1,\cdots,\Lambda_{2d}\mbox{=}1/\Lambda_d$.  Putting all
ingredients together we get the {\em \Vt} for the quantization of a
Hamiltonian dynamical system in $(d+1)$ configuration dimensions, \ie\
restricted to the fixed energy shell in the $2(d+1)$ phase space:
\beq
 \tr {\Lop}^t(E) =
 \sum_p  \period{p} \sum_{r=1}^\infty
      \prod_{i=1}^d
\frac{
      \prpgtr{t-r\period{p}} 
      \E^{({\I\over \hbar}  (S_p-ET_p) - \I \pi {\maslov{p} / 2}) r } 
     }{
	    |\Lambda_{p,i}|^{r/2}
            (1 -{ 1 / \Lambda_{p,i}^{r} })^2 
            (1 -{ 1 / \Lambda_{p,i}^{2r} }) 
     }
 \enspace .
\ee{the_Tr_Lqm}
Here $\period{p}(E) = \oint dt$ is the $p$-cycle period, 
$ 
 S_{p}(E) = \oint p dq 
$ 
the cycle action evaluated along the periodic orbit on the energy
surface $H = E$, $\maslov{p}$ the Maslov index, and $\Lambda_{p,1},
\Lambda_{p,2}, \cdots, \Lambda_{p,d}$ are the $d$ expanding eigenvalues of
the transverse \jacobianM\ of the flow belonging to the $p$-cycle.  
The period is related to the
action through 
$
 \period{p}(E) = \frac{\partial}{\partial E} S_{p}(E)
$.  
The associated \Vd\ 
is given by 
\beq
 \Zqc(E)\,=\,\exp\left\{-\sum_{p,r}
        {1 \over r}
	\prod_{i=1}^d
      \frac{ |\Lambda_{p,i}|^{-r/2}
            \E^{ \frac{\I}{\hbar} S_p(E) r 
                       - \I \pi {\maslov{p} \over 2} r}
       			    }
           {(1-1/\Lambda_{p,i}^r)^2 (1-1/\Lambda_{p,i}^{2r})}
\right\}
\label{Quasi-det}
\eeq
(see for ex. \cite{QCcourse} for the trace$\,\leftrightarrow\,$zeta
functions relationship).  This {\em \Vd} is our main result.  The
zeros of $\Zqc(E)$ yield the spectrum of the ``quasiclassical''
evolution operator.

\subsection{The Semiclassical Zeta Function}
\label{s-gutz-trace}

The formulae derived above differ from those of the semiclassical
periodic orbit theory for hyperbolic flows as originally developed by
Gutzwiller (1971) in terms of traces of the Van Vleck semiclassical
Green's functions.
The {\em semiclassical \Gt} has topologically the same structure as
the \Vt\ \refeq{the_Tr_Lqm}:
\beq
\tr G(E)
             = \overline{g}(E) + \frac{1}{\I\hbar}
                 \sum_p \period{p} \sum_{r=1}^{\infty}
        {	\E^{ \frac{\I}{\hbar} S_p(E) r 
                     - \I \pi {\maslov{p} \over 2} r}
              \over |\det \left( {\bf 1} 
                     - {\bf J}_p^r \right)|^{1 \over 2} }
\enspace .
\ee{tr-Gutz}
The \Gt\ differs from the \Vt\ in two aspects. One is the volume term
$\overline{g}(E)$ in \refeq{tr-Gutz} which is a missing from our
version of the classical trace formula. While an overall pre-factor
does not affect the location of zeros of the determinants, it plays a
role in relations such as the zeta function functional equations of
\cite{BK90}.  The other difference is that the quantum kernel leads to
a square root of the cycle \jacobian\ $1/\sqrt{\det(1- {\bf J}_p)}$, a
reflection of the relation probability~=~(amplitude)$^2$. This
difference does not effect the leading eigenvalues (which coincide for
the semi- and quasiclassical quantizations), but has a dramatic effect
on the convergence of respective zeta functions.

The precise relation between the \qS s and the \Vd s is
given in \refappe{s_GV_vs_quasi}.

In the remainder of the paper we shall investigate the relative merits
of the quasiclassical quantization compared to the Gutzwiller
semiclassics and the exact quantum mechanics.

\section{Numerical Convergence of Cycle Expansions
and Extraneous Eigenvalues}
\label{s_num_extran}

A \threedisk\ is one of the simplest classically completely chaotic
scattering systems and provides a convenient numerical laboratory for
testing both the ideas about chaotic dynamics and for computing exact
quantum mechanical spectra, see \cite{Eck87}, Gaspard and Rice
(1989a)-(1989c), \cite{CviEck89}. The
\threedisk\ consists of a free point particle moving in the 
two-dimensional plane
and scattering specularly off three identical disks of radius $a$
centered at the corners of an equilateral triangle of side length $R$.
The discrete $C_{3v}$ symmetry reduces the dynamics to motion in a
fundamental domain, and the spectroscopy to irreducible subspaces
$A_1$, $A_2$ and $E$. All our computations are performed for the fully
symmetric subspace $A_1$.

In this section we address the following question: which of the three
approximate quantization zeta functions is the best in predicting the
exact quantum mechanical scattering resonances
\begin{description}
\item[\ (a)\ ] the \qS\ of \cite{gut84} and \cite{voros88}
\beq
  \Zqm(z;k) = \exp\left\{ - \sum_{p} \sum_{r=1}^{\infty} \frac{1}{r}
 \frac{z^{r n_p}\, t_p^r }{1-{1}/{\Lambda_p^r}} \right\} 
           = \prod_p \prod_{j=0}^{\infty} 
    \left (1 - \frac{z^{n_p} t_p}{\Lambda_p^j} \right ) 
 \label{GV_zeta}
\eeq 
\item[\ (b)\ ] the \dzeta\ of \cite{Ruelle}, the $j=0$ part of the 
\qS
\beq
  \zfct{} (z;k) = \exp\left\{ 
   - \sum_{p} \sum_{r=1}^{\infty} \frac{1}{r}
 z^{rn_p}\, t_p^r \right\} 
  = \prod_p \left(1 -z^{n_p} t_p\right)
 \label{dyn_zeta}
\eeq
\item[\ (c)\ ] or the \Vd\ \refeq{Quasi-det}
\begin{eqnarray}
  \Zqc(z;k) &=& \exp\left\{ - \sum_{p} \sum_{r=1}^{\infty} \frac{1}{r}
 \frac{z^{r n_p} \, t_p^r}{
 \left ( 1-{1}/{\Lambda_p^r} \right )^2
 \left ( 1-{1}/{\Lambda_p^{2r}}\right) } \right\} \nonumber \\
 &=&\prod_p \,\prod_{j=0}^{\infty}\,\prod_{l=0}^{\infty}\, 
    \left (1 - \frac{z^{n_p} t_p}{\Lambda_p^{j+2l} }\right )^{j+1} 
 \enspace ?
 \label{qcl_zeta}
\end{eqnarray}
\end{description}
Here
\beq
 t_p= { \E^{\I k L_p - \I \maslov{p} \pi/2}}/ {|\Lambda_p|^\half}
 \label{t_p_def}
\eeq
is the weight of the $p^{\,\rm th}$ prime cycle, $n_p$ its topological
length and $z$ a book-keeping variable for keeping track of the
topological order in cycle expansions --- the above zeta functions are
Taylor-expanded in $z$ around $z=0$ up to a given cycle expansion
order and only then $z$ is set to $z=1$ (see also \refeq{cyc_exp}
below).  $L_p$ is the length of the $p^{\,\rm th}$ cycle, $\maslov{p}$
its Maslov index together with the group theoretical weight of the
studied $C_{3v}$ representation, and $\Lambda_p$ its stability (the
expanding eigenvalue of the $p^{\,\rm th}$ \jacobianM).

\begin{figure}
\centerline{{\bf(a)} \epsfig{file=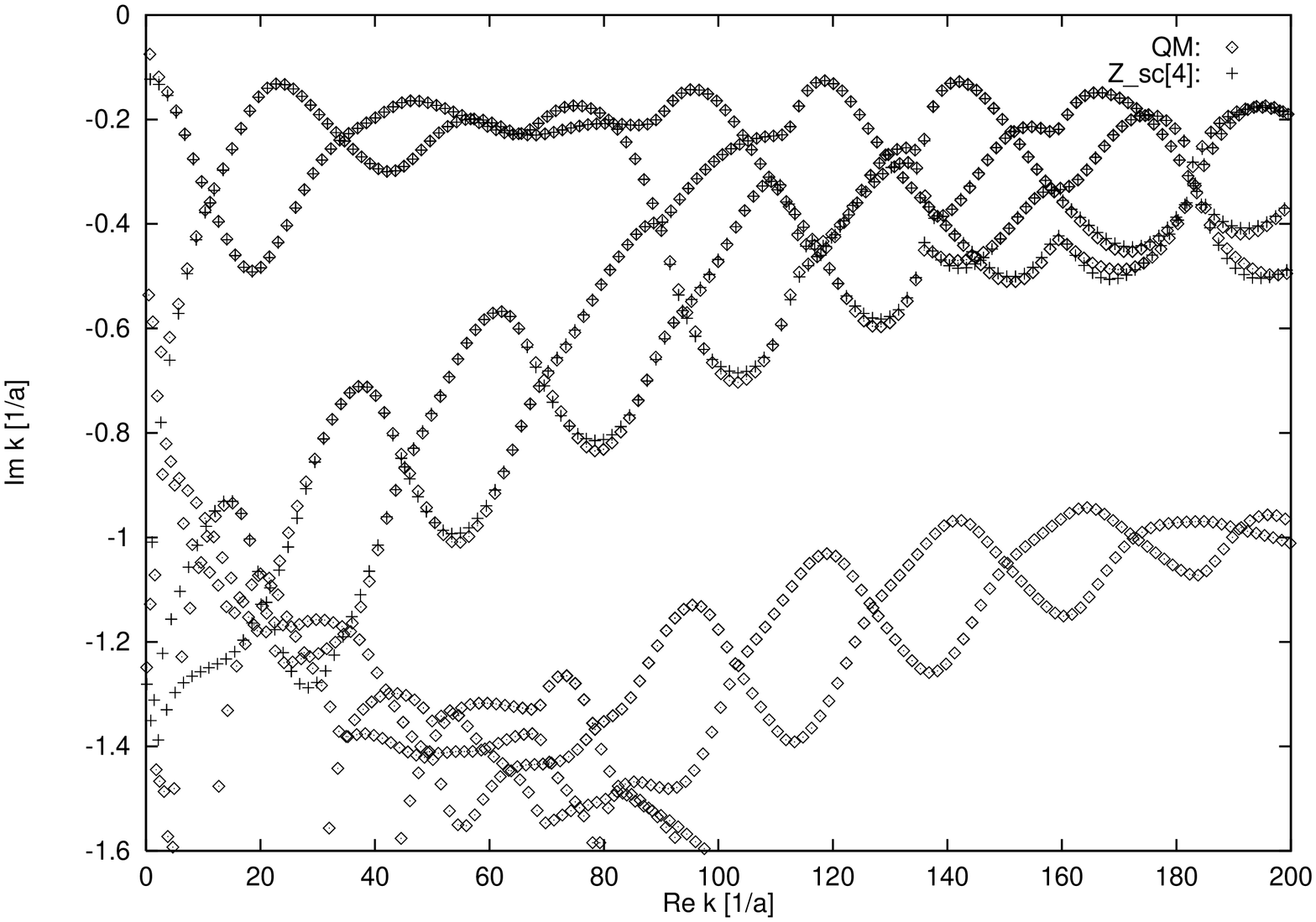,height=5.7cm,angle=-90}}

\vskip -1.5 mm
 \centerline{{\bf(b)} \epsfig{file=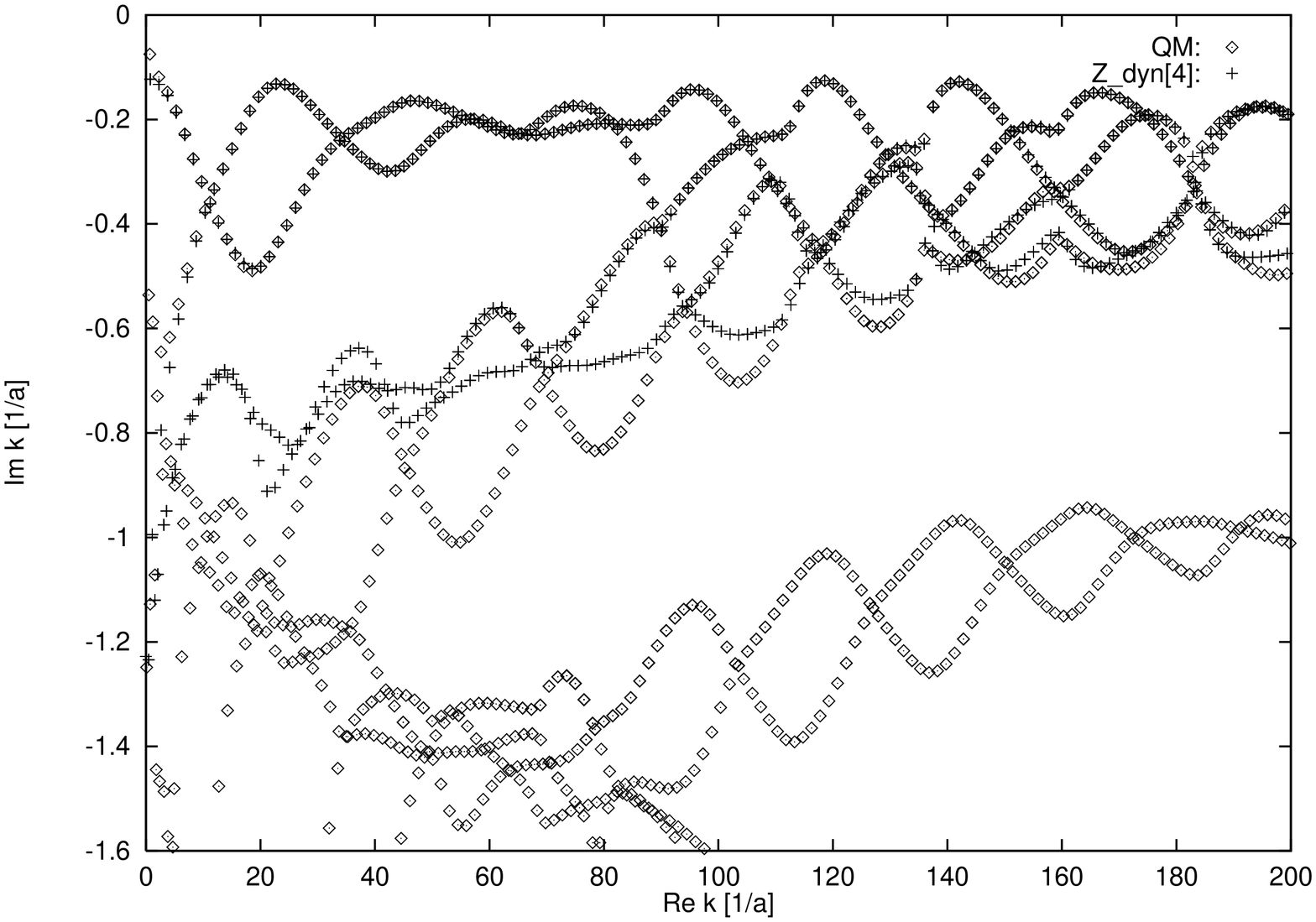,height=5.7cm,angle=-90}}

\vskip -1.5 mm
 \centerline{{\bf(c)} \epsfig{file=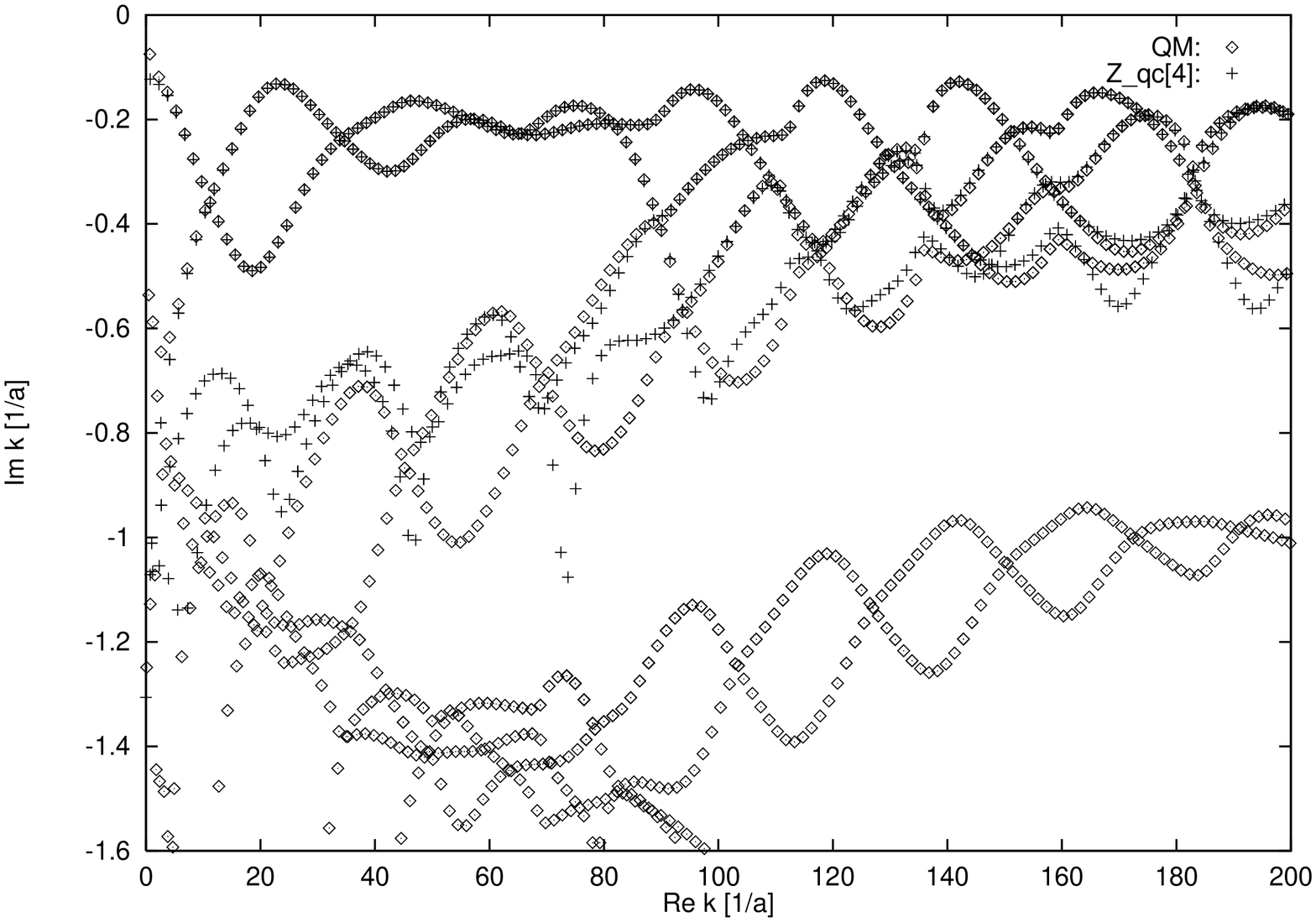,height=5.7cm,angle=-90}}
\caption[fig_e_4]{
 The $A_1$ resonances of the \threedisk\ with $R=6a$. The exact
quantum mechanical data are denoted by diamonds.  The
semiclassical ones are calculated up to $4^{\,\rm th}$ order in
the cycle expansion and are denoted by crosses. {\bf (a)}
\qS\ \refeq{GV_zeta}, {\bf (b)} \dzeta\ \refeq{dyn_zeta}, {\bf (c)} 
\Vd\ \refeq{qcl_zeta}.}
\label{fig:e_gv4}
\end{figure}

\begin{figure}
\centerline{{\bf(a)} \epsfig{file=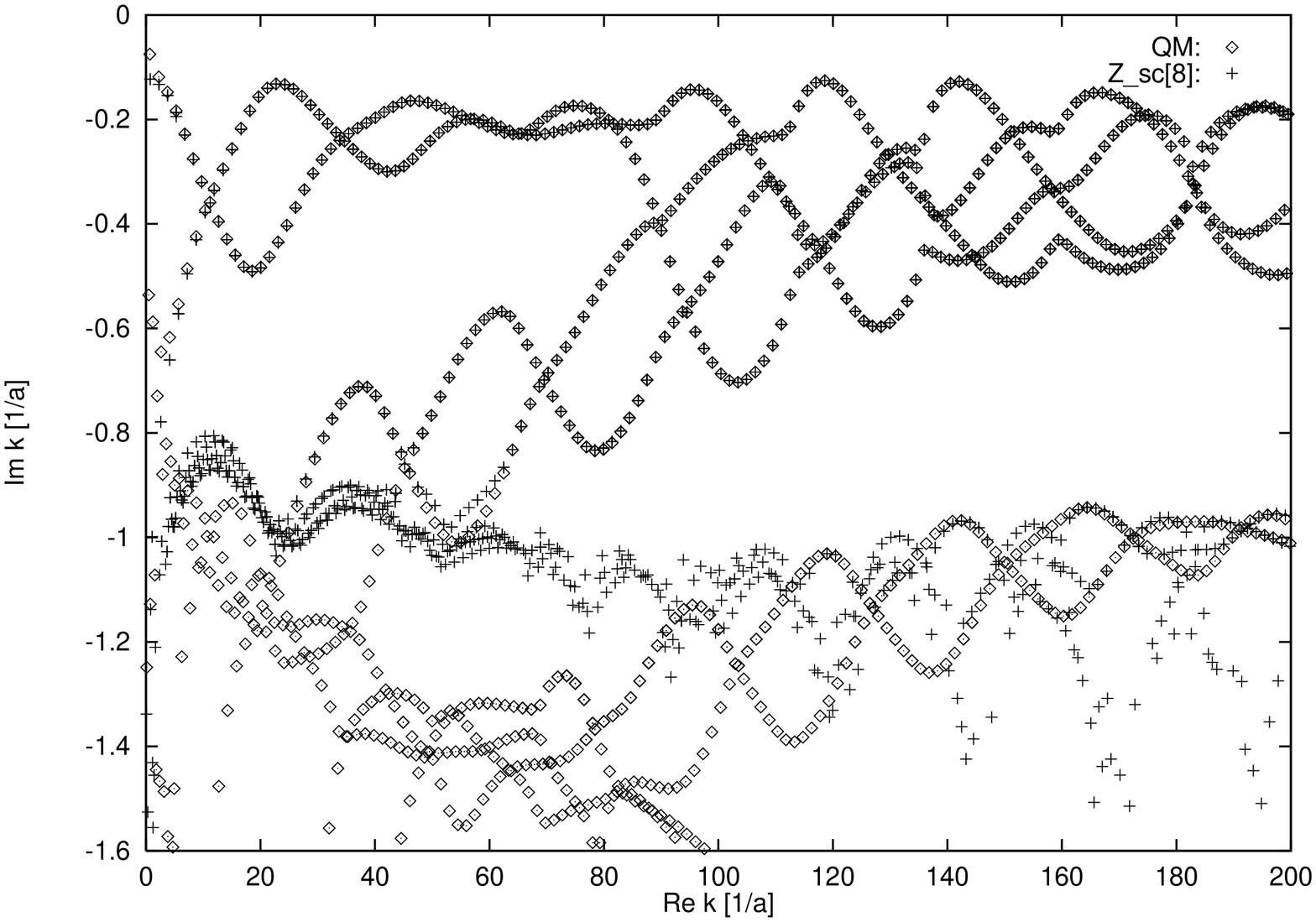,height=5.7cm,angle=-90}}

\vskip -1.5mm
 \centerline{{\bf(b)} \epsfig{file=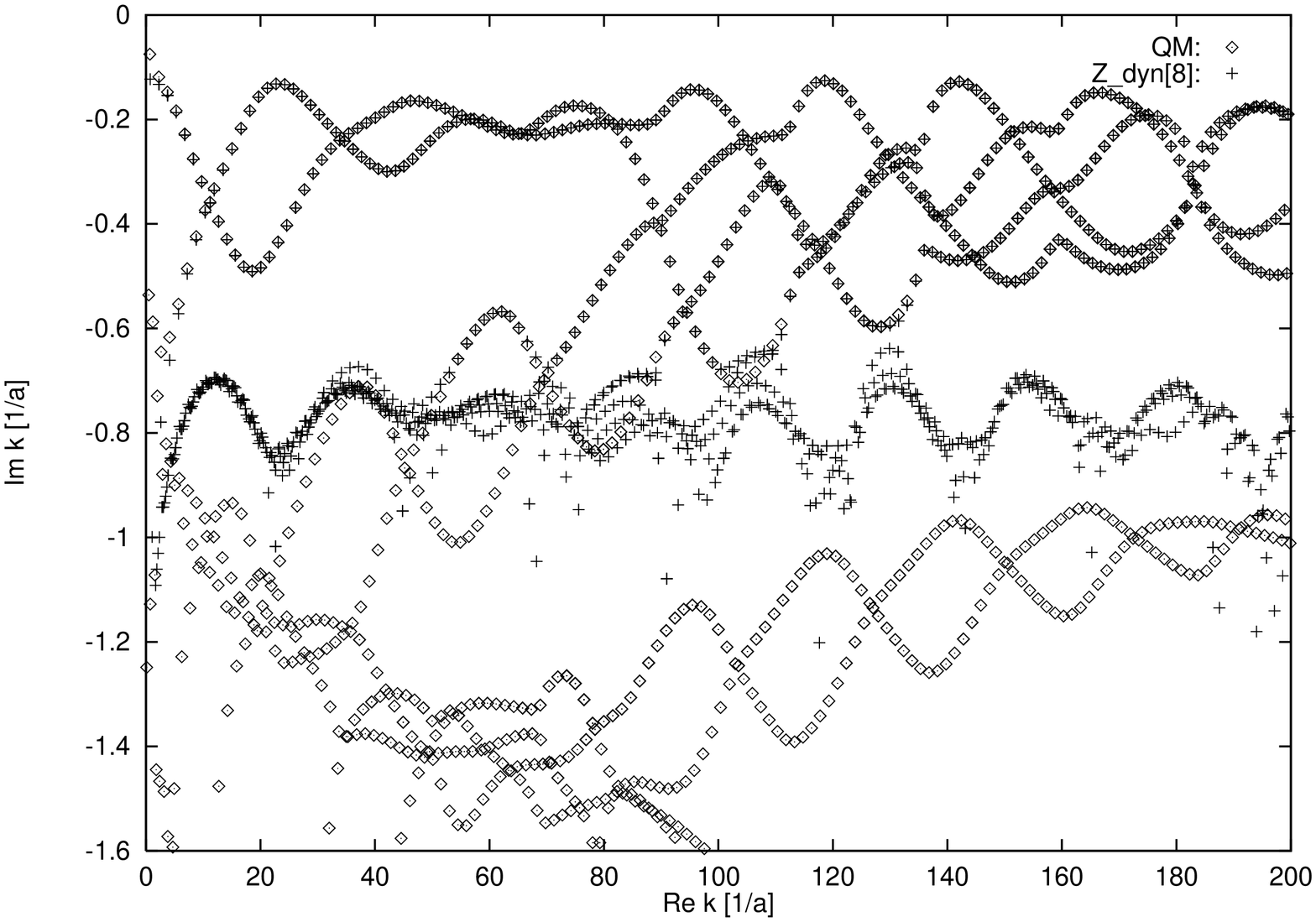,height=5.7cm,angle=-90}}

\vskip -1.5mm
 \centerline{{\bf(c)} \epsfig{file=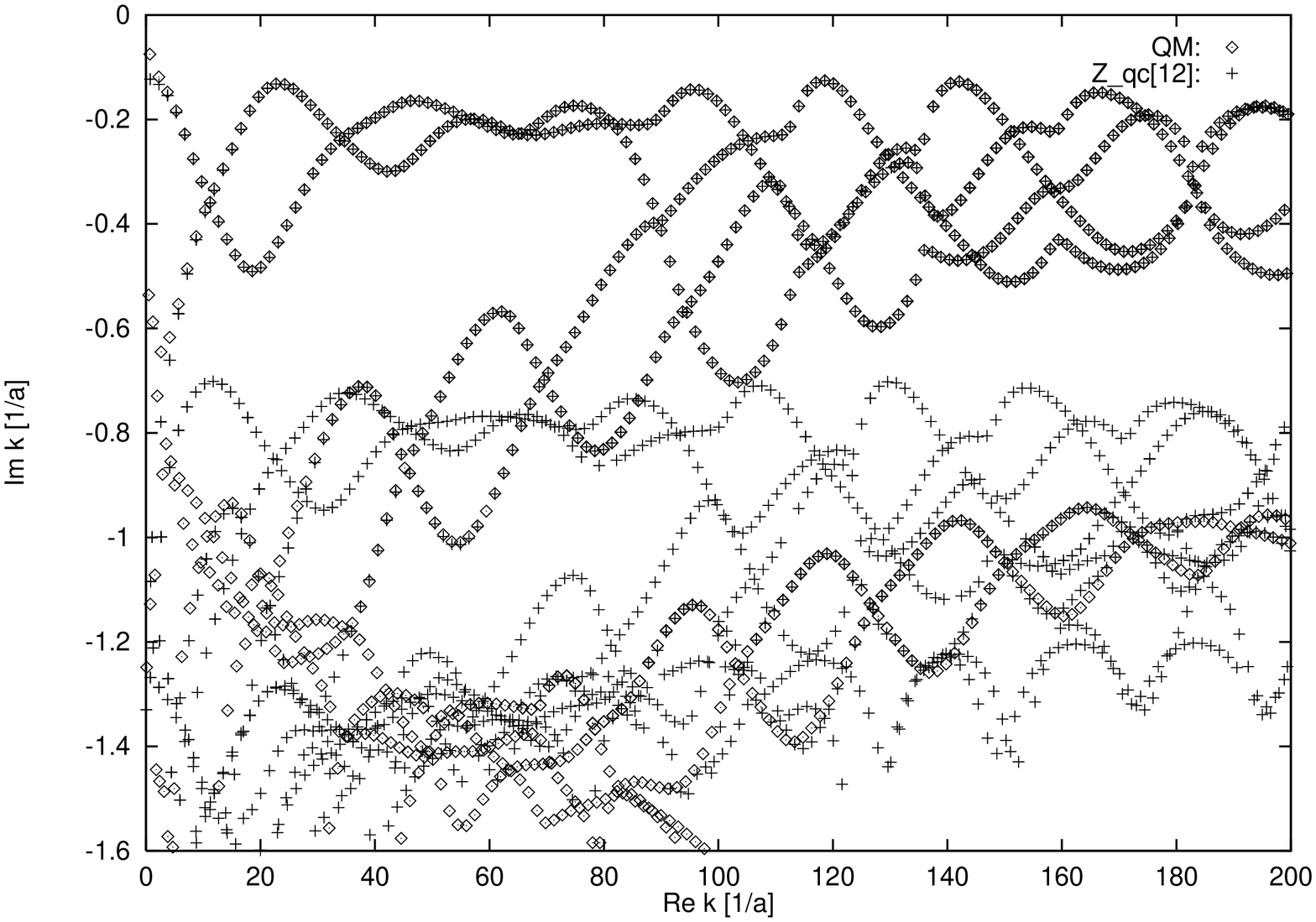,height=5.7cm,angle=-90}}

\caption[fig_e_8]{
The $A_1$ resonances of the \threedisk\ with $R=6a$. The exact
quantum mechanical data are denoted by diamonds,  the
semiclassical ones by crosses. {\bf (a)}
\qS\ \refeq{GV_zeta} up to $8^{\,\rm th}$ order in the cycle expansion, 
{\bf (b)} \dzeta\ \refeq{dyn_zeta} up to $8^{\,\rm th}$, {\bf (c)} 
\Vd\ \refeq{qcl_zeta}  up to $12^{\,\rm th}$ order.
\label{fig:e_gv8}}
\end{figure}

The results of comparing finite cycle expansion truncations of the
above zeta functions with each other and with the exact quantum
mechanical results computed with the methods outlined in
\refsect{s_semi_vs_asymp} are summarized in 
\reffigs{fig:e_gv4} and \ref{fig:e_gv8}.
Resonances are plotted as the real part of the resonance wavenumber
(resonance ``energy'') vs. the imaginary part of the wavenumber
(resonance ``width'').  We have computed several thousands of exact
quantum mechanical as well as approximate $A_1$ resonances for the
\threedisk\ with center-to-center separation $R=6a$. Further and
considerably more detailed numerical results are available from
\cite{WH95}.

Some of the features of the resonance spectra have immediate
interpretation. The mean spacing of the resonances is approximately
$2\pi/{\bar L}$, where ${\bar L}$ is the average of the lengths
$L_0$ and $L_1$ of the two shortest cycles of topological length
one. The data also exhibit various beating patterns resulting from the
interference of cycles of nearly equal lengths; e.g. the leading
beating pattern is of order $2\pi/{\Delta L}$, where $\Delta L$ is the
difference of the lengths $L_1$ and $L_0$.
 
In \reffig{fig:e_gv4} the cycle expansion includes cycles up to
topological length four. Already at this order the four leading
resonance bands are well approximated by the
\qS\ \refeq{GV_zeta} (in fact, for ${\rm Re}\, k
\lesim 75/a$ already cycles up to length two suffice
to describe the first two leading resonance bands). 
Neither the \Vd\ \refeq{qcl_zeta} nor the \dzeta\ \refeq{dyn_zeta}
perform quite as well.  The reason is that the quasiclassical as well
as the \dzeta\ predict extra resonances which are absent in the exact
quantum mechanical calculation.  The accessible resonances close to
the real axis can in this regime be parameterized by 16 measured
numbers,
\ie\ 8 cycle lengths and stabilities, together with the 8 Maslov
indices.  It turns out that the subleading bands remain completely
shielded all the way up to ${\rm Re}\, k \approx 950/a$ where they
start mixing with the four leading ones.

{In \reffigs{fig:e_gv8}a-c}\ the comparison is made up to eight,
respectively twelfth cycle expansion order.  
The border of convergence of the
\qS\ has now moved (in the plotted region) above the fifth
and sixth band of the exact quantum resonances. The \dzeta\ exhibits a
sharp accumulation line of resonances, the border of convergence
controlled by the location of the nearest pole of the 
\dzeta\,(\cite{ER92,CRR93}).  With cycles up to length 12 the \Vd\ resolves
the exact quantum fifth and sixth bands of subleading resonances, but
at the cost of many extraneous resonances, see \reffig{fig:e_gv8}(c).
At these high cycle expansion orders the
\Vd\  has convergence problems for large
negative imaginary $k$ values (especially for low values for ${\rm
Re}\, k$), in agreement with the expected large cancellations in the
cycle expansion at high cycle expansion orders, \cite{WH95}.  There is
the further caveat that the \Vd\ finds the lowest subleading
resonances just barely at the 12$^{\rm th}$ order in the cycle
expansion. Therefore cycles of larger topological length would be
needed to confirm this success.

The extraneous eigenvalues are not without a meaning; they belong to
the spectra of classical \evOper s, such as those that describe the
escape from a classical \threedisk, plotted in \cite{CRR93}. The
problem is that we now know, by comparing them to the exact quantum
mechanical spectra, that they have nothing to do with quantum
mechanics. As far as quantum mechanics is concerned, they are
``extraneous''.

Another distinctive feature of the exact quantum mechanical spectra is
the {\em diffractive} band of resonances from $k\approx(0. -\I 0.5)/a$
to $k\approx (100.-\I 1.6)/a$.
As shown by \cite{VWR94} and \cite{RVW96}, the diffractive band of
resonances can be accounted for by inclusion of creeping periodic
orbits, omitted from the calculations undertaken here.

Qualitatively, the results can be summed up as follows. The
\qS\ \refeq{GV_zeta} does well above the line of convergence
defined by the \dzeta\ \refeq{dyn_zeta}, already at very low cycle
expansion orders where the other two zeta functions still have
problems. Below this line of convergence the
\qS\ works only as an asymptotic expansion; when it
works, it works very well and very efficiently. The 
\dzeta\ does eventually as well for the leading resonances as
the semiclassical one. As experimentally these are the only resonances
accessible, one can -- for practical purposes -- limit the calculation
just to this zeta function.  The \Vd\ finds all known subleading
quantum resonances, but at a high expense: the rate of convergence is
poor compared to the \qS, as most of the information provided by
longer cycles is used to determine the extraneous resonance bands,
with no quantum counterpart. Without a quantum calculation, one could
not tell the extraneous from the real resonances.

As a by-product of this calculation we can state an empirical rule of
thumb: Each new cycle expansion or cumulant order is connected with a
new line of subleading resonances.  This rule relates the cycle
expansion truncations limit, $n\to\infty$ (where $n$ is defined below
in \refeq{cyc_exp}), and the limit ${\rm Im}\, k \to -\infty$.
Numerics supports the claim that the cycle expansion limit
$n\to\infty$ and the semiclassical limit ${\rm Re}\, k \to \infty$ do
not commute deep down in the lower complex $k$ plane, a point that we
shall return to in
\refsect{s_semi_vs_asymp}.

\subsection{Exact Versus Semiclassical Cluster Phase Shifts}
\label{s_phas_shift}

In the above we compared the exact and semiclassical resonances of the
\threedisk\ in the $A_1$ representation.  As the deviations are most
pronounced for the subleading resonances which are shielded by the
leading ones, one could argue that experimentally it does not matter
which of the three zeta functions are used to describe the measured
data, as all three give the same predictions for the leading
resonances.

Nevertheless, as we shall now show, the three approximate
quantizations can be told apart (\cite{Wi95}), even experimentally.

The exact and semiclassical expressions for the determinant of the
$S$-matrix for the non-overlapping
\threedisk\ are given by 
\begin{eqnarray}
  \det\, {\bf S}^{(3)}(k) 
   &=& \left (  \det\,{\bf S}^{(1)} (ka)\right )^3 
     \frac{ \det\, { {\bf M}_{A1} (k^\ast) }^\dagger}
          { \det\,  {\bf M}_{A1} (k) }\,
     \frac{ \det\,  { {\bf M}_{A2} (k^\ast) }^\dagger}
          { \det\,  {\bf M}_{A2} (k) }\,
   \frac{\left (  {\det\, {\bf M}_{E} (k^\ast)}^\dagger\right )^2 }
          { \left( \det\,  {\bf M}_{E} (k)\right )^2 }    \nnu \\
  &\semiclass &
         \left ( \E^{-\I \pi N(k)} \right )^{6}
      \, \left( \frac{  {{Z}_{\rm 1\mbox{-}disk(l) }(k^\ast)}^\ast}
            {{{Z}_{\rm 1\mbox{-}disk(l)} (k)}}\, 
        \frac{{ {Z}_{\rm 1\mbox{-}disk(r)}
              (k^\ast)}^\ast}
            {{{Z}_{\rm 1\mbox{-}disk(r)} (k)}}\right )^3 
 \times \nnu \\
  && \qquad \qquad \qquad\times
            \frac{{ {Z}_{A1} (k^\ast)}^\ast  }
                     {{Z}_{A1}(k)}\,
               \frac{ {{Z}_{A2} (k^\ast)}^\ast }
                {{Z}_{A2}(k)}
           \,
              \frac{ { {{Z}_{E} (k^\ast)}^\ast  }^2}
                      {{{Z}_{E}(k)}^2}  \enspace .
 \label{gen_sym}
\end{eqnarray}
(See \cite{WH95} for details and notation.)  For the $A_1$
representation of the \threedisk\ the quantum mechanical kernels and
the \qS s \refeq{GV_zeta} are related by
\begin{eqnarray}
   \frac{ \det\, { {\bf M}_{A1} (k^\ast) }^\dagger}
          { \det\,  {\bf M}_{A1} (k) }
\semiclass 
 \frac{{ {Z}_{A1} (k^\ast)}^\ast  }
                     {{Z}_{A1}(k)} 
    \label{link}
\end{eqnarray}
Both sides of \refeq{gen_sym} and \refeq{link} respect unitarity, and
if the wave number $k$ is real, both sides can be written as $\exp\{
\I 2 \eta(k) \}$ with a real {\em phase shift} $\eta(k)$.  We define
the total phase shift for the coherent part of the 3-disk scattering
problem (here always understood in the $A_1$ representation) for the
exact quantum mechanics as well as for the three approximate
quantizations by:
\begin{eqnarray}
    \E^{2\I \eta_{\rm qm}(k)} &:=&    
\frac{ \det\, { {\bf M} (k^\ast) }^\dagger}
          { \det\,  {\bf M} (k) }  \qquad  
      \E^{2\I \eta_{\rm sc}(k)}\,:=\,
 \frac{{ \Zqm (k^\ast)}^\ast  }
                     {\Zqm(k)} \continue 
      \E^{2\I \eta_{\rm dyn}(k)}&:=&
 \frac{{  \zfct{} (k^\ast)}^\ast  }
                     { \zfct{}(k)}  \qquad \ \ \, 
      \E^{2\I \eta_{\rm qc}(k)}\,:=\,
 \frac{{ \Zqc (k^\ast)}^\ast  }
                     {\Zqc(k)} 
 \label{eta_qcl}\enspace .
\end{eqnarray}
This phase shift definition should be compared with the cluster phase
shift given in section~4 of \cite{LS72}.  The important point here is
that the coherent or cluster phase shift of $\det\,{\bf S}(k)$ is in
principle experimentally accessible: one just has to construct the
elastic scattering amplitude from the measured cross sections, and
subtract the single disk contributions.

\begin{figure}
\centerline{ \epsfig{file=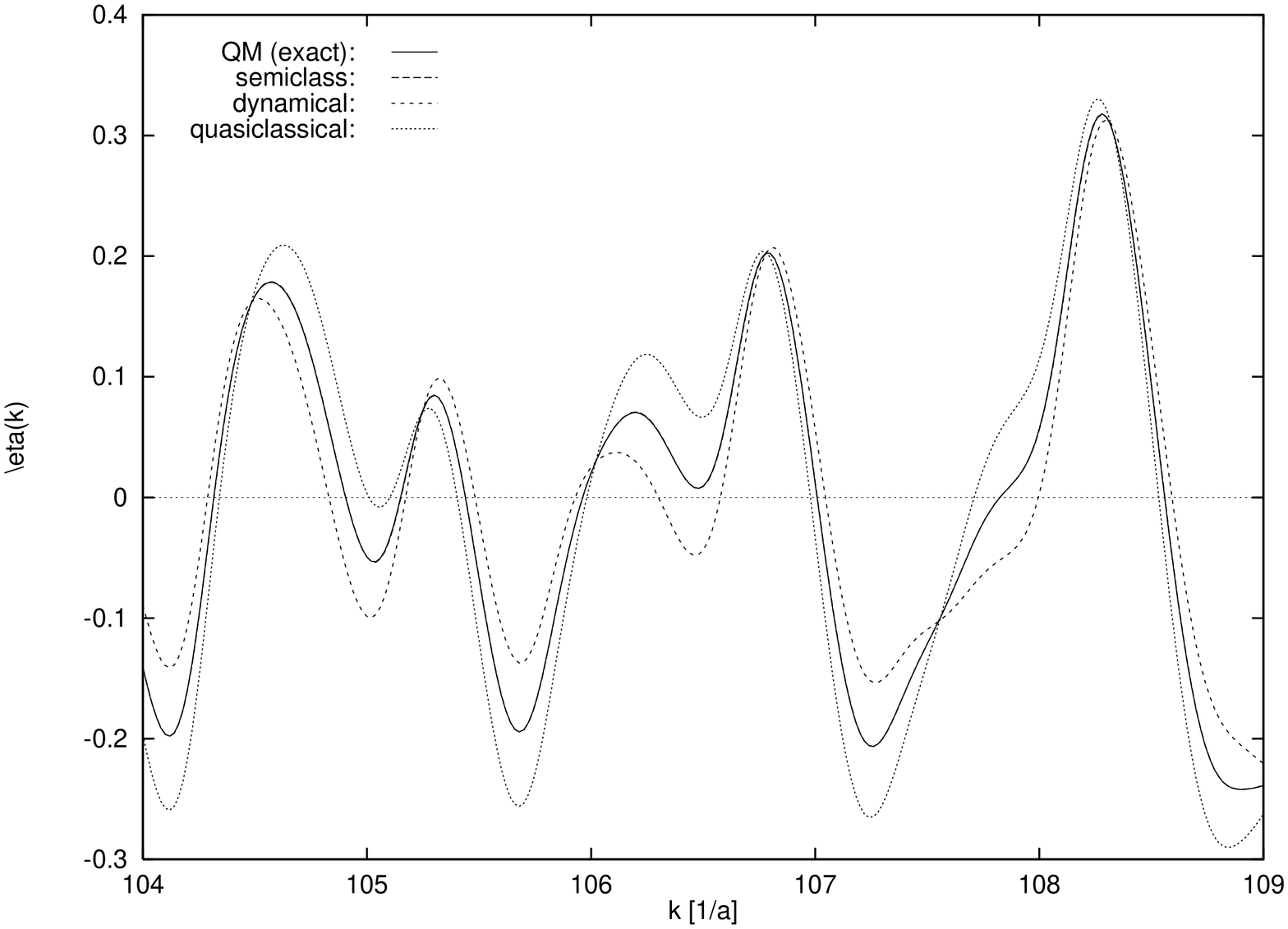,height=8cm,angle=-90}}
\caption[fig_ph_gv]{
The coherent cluster phase shifts of the 3-disk scattering system in
the $A_1$ representation with $R=6a$. The exact quantum mechanical
data compared to the predictions of the
\qS\ \refeq{GV_zeta}, the \dzeta\ \refeq{dyn_zeta} and 
the \Vd\ \refeq{qcl_zeta} calculated up to
$12^{\,\rm th}$ order in the cycle expansion. The \qS\ and the
exact quantum mechanical data coincide  within the resolution
of the plot.
\label{fig:cluster}
 }
\end{figure}

So, $\eta_{\rm qm}(k)$ is a ``measurable'' quantity, useful to us as a
different method for discriminating between the various zeta
functions. An example is given in \reffig{fig:cluster} where the zeta
functions in the numerators as well as in the denominators in
\refeq{eta_qcl} have been expanded up to cycles of topological length
12.  The phase shifts are compared in the window $ 104/a \leq k \leq
109/a$, a typical window sufficiently narrow to resolve the rapid
oscillations, with $k$ sufficiently big that the diffraction effects
are unimportant.
%
%
The performance of the original \qS\ is again the best.  We stress
that in contrast to the subleading resonances studied in
\refsect{s_num_extran} (which are completely shielded from
experimental detection by the leading resonances), phase shifts are
{\em hard} data, in principle extractable from measured cross
sections.

In conclusion: 
One can tell the three candidate zeta functions apart
even experimentally.  We have again confirmed that the \qS\ is the
best.

\section{Semiclassics Versus Asymptotic $\hbar$ Expansion}
\label{s_semi_vs_asymp}

So far we have tested various approximate quantization proposals
against each other and against exact quantum mechanics.  Now we turn
to a deeper question: how seriously should we take these cycle
expansions in the first place?  We will show here, following
\cite{Wi96}, that the \qS\ is approximating its quantum mechanical
counterpart, the ``characteristic KKR determinant''
(\cite{KKR,LS72,Berry_KKR}) as an asymptotic series and therefore
makes sense only as a truncated series.

Let $\det\, {\bf M}(k)=\det({\bf 1}+ {\bf A}(k))$ be the
characteristic KKR determinant of the
\threedisk\ in the $A_1$ representation, where the 
pertinent kernel ${\bf A}(k)$ expressed in 
the angular momentum basis relative to the half-disk in
the fundamental domain reads (see \cite{gr_qm})
\begin{eqnarray}
 {\bf A}(k)_{m,m'} &=& d(m) d(m') \frac{\Jb{m}{a}}{\Ho{m'}{a}}\left\{
 \cos\left(\frac{\pi}{6}(5m-m')\right) \Ho{m-m'}{R} \right.\nnu \\
 &&\qquad \mbox{}\left.
 +(-1)^{m'}\cos\left(\frac{\pi}{6}(5m+m')\right) \Ho{m+m'}{R} 
\right\} 
\end{eqnarray}
with $ 0\leq m,m' < \infty$ and 
\begin{eqnarray}
 d(m):=\left \{ \begin{array}{ccc} \sqrt{2} & {\rm for} & m>0 \\
                                   1        & {\rm for} & m=0 
            \enspace . 
              \end{array} \right.
 \nnu
\end{eqnarray}
Let $Q_m(k)$ denote the $m^{\,\rm th}$ cumulant of $\det\, {\bf
M}(k)$,
\ie\ the coefficient of $z^m$ in the Taylor expansion of
$\det({\bf 1}+z {\bf A}(k))$. $Q_m(k)$ satisfies the Plemelj-Smithies
recursion relation (\cite{WH95})
\begin{eqnarray}
 Q_m (k) &=& \frac{1}{m}\sum_{j=1}^{m} (-1)^{j+1} Q_{m-j}(k)\, 
   {\rm Tr}({\bf A}^j(k)) \quad {\rm for}\ m \geq 1 \nonumber \\ 
 Q_0 (k) &\equiv & 1 \enspace , \nonumber
\end{eqnarray} 
where ${\rm Tr}({\bf A}^j (k))$ is the trace of the $j^{\, {\rm th}}$
power of the kernel ${\bf A}(k)_{m,m'}$ evaluated in the angular
momentum basis, $\{|m\rangle\}$, relative to the half-disk in the
fundamental domain.

\begin{figure}
\centerline{ 
\epsfig{file=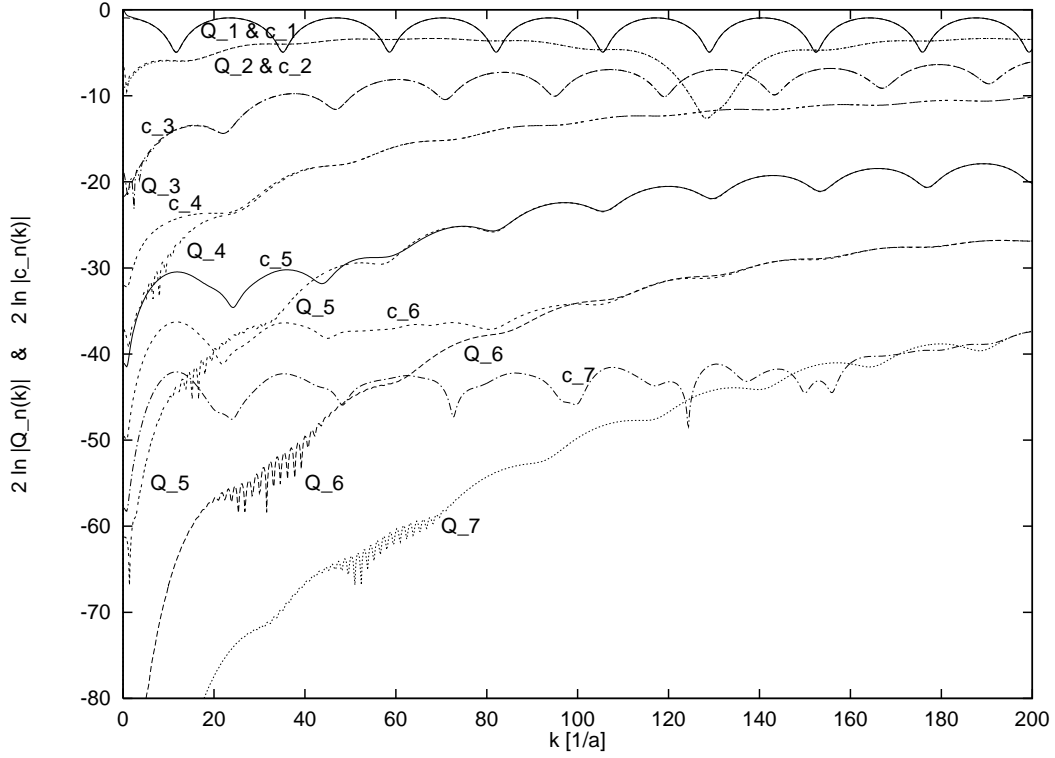,height=10cm,angle=-90}}
\caption[fig_cum_mod]{
Comparison of the absolute values of the first seven quantum
mechanical cumulant terms, $|Q_n(k)|^2$, with the corresponding
semiclassical cycle expansion terms, $|c_n(k)|^2$, of the
\qS\ \refeq{GV_zeta} evaluated on the real wave number axis $k$. 
Note that the deviations between quantum mechanics and semiclassics
decrease with increasing ${\rm Re}\, k$, but increase with increasing
cycle expansion order $n$.  The value of ${\rm Re}\, k$ where the
quantum mechanical and semiclassical curves join is approximately
given by ${\rm Re}\, k a \sim 2^{n+1}$ where $n$ is the order of the
cumulant/cycle expansion term and $a$ is the radius of the disk.
$A_1$ \threedisk\ with center-to-center separation $R=6a$.
\label{fig_cum_mod} }
\end{figure}

The semiclassical analog of the characteristic determinant $\det({\bf
1}+z {\bf A}(k))$ is the \qS\ \refeq{GV_zeta}.  More precisely, the
cycle expansion of the \qS\ truncated at the topological order $n$ is
the semiclassical analog of the quantum cumulant expansion of
$\det({\bf 1}+z {\bf A}(k))$ truncated at the same order.
Thus $ c_m(k)$, the
corresponding semiclassical $m^{\,{\rm th}}$ order cycle expansion
term of $\Zqm(k)$, is constructed from the semiclassical
equivalent of the Plemelj-Smithies recursion relation: 
\begin{eqnarray} 
 c_m (k) &=& \frac{1}{m}\sum_{j=1}^{m} 
 (-1)^{j+m+1} c_{m-j}(k) \sum_{p,r} n_p\,
 \frac{\delta_{n_pr,j} {t_p}^r}{1-1/\Lambda_p^r} 
 \qquad {\rm for}\ m \geq 1 
 \label{cm_semicl} \\
 c_0 (k) &\equiv & 1  
 \nnu 
 \enspace ,
\end{eqnarray}    
with $t_p$ defined in \refeq{qcl_zeta}.  The cycle expansion
(\cite{cycprl}) follows from the semiclassical limit
\begin{eqnarray}
  {\rm Tr} ({\bf A}^j(k) ) \semiclass  (-1)^j
 \sum_{p,r} n_p\, \frac{\delta_{n_pr,j} {t_p}^r}{1- 1/\Lambda_p^r} 
   \ +  
   \mbox{\small diffractive creeping orbits}\enspace . 
 \label{conjecture}
\end{eqnarray} 
In summary, the $n^{\,\rm th}$ order truncated 
cumulant and cycle expansions are given by
\beq
      \det\, {\bf M}(k)|_n = \sum_{m=0}^{n}  Q_m (k) 
	\,, \qquad 
       \Zqm(k)|_n   =  \sum_{m=0}^{n}  c_m(k)        
\ee{cyc_exp}
where the notation $\cdots|_n$ indicates that the corresponding
determinant or zeta function has been truncated at cumulant/cycle
expansion order $n$.  The following facts are known:
\begin{enumerate}   
\item
The cumulant sum
\[  
     \lim_{n\to \infty}   \det\, {\bf M}(k)|_n =   \lim_{n\to \infty} 
 \sum_{m=0}^{n} Q_m(k) = \det\,{\bf M}(k)
\]
is absolutely convergent, $ \sum | Q_m(k) | < \infty \enspace $,
because of the trace class property of 
${\bf A}(k)\equiv {\bf M}(k)-{\bf 1}$ 
for non-overlapping, non-touching
\ndisk s (\cite{WH95}).
\item
The semiclassical cycle expansion sum converges above an accumulation
line (which runs below and approximately parallel to the real wave
number axis, see \reffig{fig:e_gv8}(a)) given by the leading poles of
the leading \dzeta,
$\zfct{}(k)$, or the leading zeros of the subleading zeta function,
$\zfct{1}(k)$ (\cite{ER92,CRR93,CV93}).
\item
The {\em truncated} semiclassical cycle expansion sum $\Zqm(k)|_n$ can
approximate the quantum mechanical result as an asymptotic series even
below the \qS\ border of convergence, \cite{WH95}.
\end{enumerate}
We have checked numerically that the following formulae relate the
$m^{\,\rm th}$ cumulants and cycle expansion terms on the real
$k$-axis with the corresponding quantities inside the complex $k$
plane --- at least as long as the condition $| {\rm Im}\, k| \ll |{\rm
Re}\,k |$ is satisfied: for the quantum mechanical cumulants of order
$m$ we have the approximate leading order relation (under assumption
that the diffraction effects are negligible)
\beq
    Q_m({\rm Re}\, k + \I {\rm Im}\, k) \sim Q_m({\rm Re}\, k) \,
   \E^{- m \bar{L} {\rm Im}\, k} \enspace .  
\ee{Qscale} 
$\bar{L}\approx R-2a$ is the average length of the cycles of
topological length one.  We have also checked numerically that the
corresponding relation for the semiclassical cycle expansion terms
of order $m$ is also approximately valid:
\begin{eqnarray}    
    c_m({\rm Re}\, k + \I {\rm Im}\, k) \sim   c_m({\rm Re}\, k) 
    \, \E^{ - m \bar{L}  {\rm Im}\, k} \enspace .
  \label{Cscale}
\end{eqnarray}
Furthermore, on the basis of \reffig{fig_cum_mod} 
%
%
we conjecture that for arbitrary values of the center-to-center
separation $R$ of the non-overlapping \threedisk\ ($R>2a$) the
following relations hold on the real wave number axis ($k$ real):
\begin{eqnarray}
   c_m(k) \approx Q_m(k) \quad {\rm with}\ \ 
 1 \gg |c_m(k)|    \qquad  
  {\rm if}\  \ ka  \gesim\,  2^{m-1} \frac{\bar L}{a}  \enspace ,
  \label{asymp} \\
     1 \gg |c_m(k)|  \gg | Q_m(k) | \qquad       
 {\rm if}\ \  ka  \lesim\,   
   2^{m-1} \frac{\bar L}{a} \enspace .  
   \label{non_asymp}
\end{eqnarray}

\subsection{The Meaning of It All}

Where is the boundary $ka \approx 2^{m-1} {\bar L}/{a}$ coming
from?

This boundary follows from a combination of the uncertainty principle
with ray optics and the non-vanishing value for the topological
entropy of the \threedisk. When the wave number $k$ is fixed, quantum
mechanics can only resolve the classical repelling set up to the
critical topological order $n$ given by \refeq{asymp}. The quantum
wave packet which explores the repelling set has to disentangle $2^n$
different sections of size $d \sim a / 2^n$ on the ``visible'' part of
the disk surface (which is of order $a$) between any two successive
disk collisions.  Successive collisions are separated spatially by the
mean flight length $\bar L$, and the flux spreads with a factor $\bar
L/a$.  In other words, the uncertainty principle bounds the maximal
sensible truncation in the cycle expansion order by the highest
quantum resolution attainable for a given wavenumber $k$.

The upper limit $n$ for which $c_m(k)$ with $ m \leq n$ approximates
$Q_m(k)$ is increasing with increasing ${\rm Re}\, k$.  For $n >
m({\rm Re}\, ka)$, defined in \refeq{non_asymp}, the cycle expansion
terms and cumulant terms deviate so much from each other, that beyond
this order 
the contributions of longer cycle expansions have
nothing to do with quantum mechanics.  The fact that $\Zqm(k)|_n$ --
even in its convergence regime -- is a good approximation to quantum
mechanics {\em only} up to a finite $n$ is usually not noticed, 
as the terms
in \refeq{non_asymp} are 
exponentially small on or close to the real axis
and sum therefore to a tiny quantity. In other words, for $n> m({\rm
Re}\, ka)$ and close to the real $k$ axis, the absolute error
$|c_n(k)-Q_n(k)|$ is still small, the relative error $|c_n(k)/Q_n(k)|$
on the other hand is tremendous.  With increasing negative ${\rm Im}\, k$,
however, using the scaling rules \refeq{Qscale} and \refeq{Cscale},
the deviations \refeq{non_asymp} are blown up, such that the relative
errors $|c_n(k)/Q_n(k)|$ eventually become visible as absolute errors
$|c_n(k)-Q_n(k)|$ (see e.g. the resonance calculation of \cite{WH95}).
For ${\rm Im}\, k$ above the boundary of convergence these errors
still sum up to a finite quantity which might, however, not be
negligible any longer.  Below the convergence line these errors sum up
to infinity.

So, the value of ${\rm Im} k$ where --- for a given $n$ --- the
$\Zqm(k)|_n$ sum deviates from $\det\, {\bf M}(k)|_n$ is governed by
the real part of $k$ and the scaling rules \refeq{Qscale} and
\refeq{Cscale}.  It has nothing to do with the boundary of convergence
of $\Zqm(k)$, as a good approximation
is given by the {\em finite} sum of terms satisfying
\refeq{asymp}. Therefore, the truncated semiclassical expansion can
describe the quantum mechanical resonance data even {\em below} the
line of convergence of the infinite cycle expansion series, as we have
already noted in \refsect{s_num_extran}.

On the other hand, the boundary line of the convergence regime of the
semiclassical expansion is governed by $c_m(k)$, $m\to\infty$, terms
which have nothing to do with the quantum analog $Q_m(k)$, \ie\ solely
by terms of type
\refeq{non_asymp}.
The reason is that the convergence property of an infinite sum is
governed by the {\em infinite} tail and not by the first few terms.
Whether a semiclassical expansion converges or not is a separate issue
from the question whether the quantum mechanical data are described
well or not.  The {\em convergence property} of a \qS s on the one
hand and the {\em approximate description of quantum mechanics} by
these zeta functions are therefore two {different issues}.  It could
happen that a zeta function is convergent, but not equivalent to
quantum mechanics, as we have seen was the case with the extraneous
resonances in the quasiclassical calculation. Or that it is not
convergent in general, but its finite truncations nevertheless
approximate well quantum mechanics, as is the case for the
Gutzwiller-Voros \qS \refeq{GV_zeta}.

We conclude that the exponential rise of the number of cycles with
increasing cycle expansion order $n$ is the physical reason for the
breakdown of the cycle expansion of the \qS \refeq{GV_zeta} with
respect to the exact quantum mechanical cumulant expansion.
                        
\section*{Summary and Conclusions}

In conclusion, we have constructed a classical
\evOper\ for the quasiclassical wave function evolution,
and derived the corresponding trace and determinant formulae for
periodic orbit quasiclassical quantization of chaotic dynamical
systems.

Improved analyticity has been very useful in sorting out the relative
importance of the semiclassical, diffraction and quantum
contributions.  However, one hope for consequence of the
superexponential convergence of the cycle expansions of the new \Fd\
was that they would converge faster with the maximal cycle length
truncation than the more familiar Gutzwiller-Voros and Ruelle type
zeta functions.  As is shown here, this is not the case. Improved
analyticity comes at cost; extraneous eigenvalues are purely classical
and do not belong to the quantum spectrum, but their presence degrades
significantly the convergence of the cycle expansions.

The analysis sheds new light on the differences between the classical
and semiclassical spectra; in particular, we have made explicit for
the case of $n$-disk repellers the quantum limitations on the phase
space resolution by classical orbits, in the spirit of \cite{Bog92}
analysis of the finite resolution of phase space for the bound
systems.

In spite of its laggard performance as a putative competitor to the
semiclassical quantization, the mere fact that there exists an
alternative ``quasiclassical'' quantization that follows directly from
the Schr\"o\-din\-ger equation without recourse to path integrals and
saddle points is of intellectual interest.  It is still possible that
a more ingeniously constructed ``classical'' evolution operator would
also perform better than the
\qS\ in practice.

\vskip 1cm
\noindent
{\bf Acknowledgements.}
It was Dieter Wintgen's success in applying cycle expansions
to atomic physics that has motivated the above ``in depth'' study
of various semiclassical quantizations.
We are grateful to Per E. Rosenqvist for a long and fruitful collaboration,
and in particular for the \threedisk\ data set used in the
numerical experiments of this paper. We have striven to
meet Gregor Tanner's exacting standards. G.V. thanks the grant of the
Hungarian Ministry of Culture and Education (MKM 337) and of the
OTKA ( F019266/F17166/T17493 ). 
G.V. 
and A.W. 
acknowledge the support of
the International Relations Offices of Germany and Hungary. 
A.W. is grateful to Michael Henseler for discussions and
would like to thank the Center for Chaos and Turbulence Studies at
the Niels Bohr Institute for hospitality during his visits to Copenhagen.
\appendix
\section{Calculation of Trace ${\bf M}$}
\label{App_Jacob}

In this appendix we calculate the trace \refeq{Tr_La2}.
The equations of motion 
for a time independent Hamiltonian \refeq{Ham_eqs} 
can be written as
\beq
\dot{\pSpace}_m = {\bf \omega}_{mn} 
	\frac{\partial H}{\partial \pSpace_n}\enspace , \qquad 
\quad {\bf\omega} = \MatrixII{0}{\bf I}{\bf -I}{0}
	\,,\qquad
	m,n=1,2,\dots,2d
\enspace ,
\ee{symplect}
where $\pSpace = [q,p]$ is a phase space vector,    ${\bf I} =$ 
[$d\times d$] unit matrix, and $\omega$ the [$2d\times 2d$] symplectic
form ${\bf \omega}_{mn} = - {\bf \omega}_{nm}$, ${\bf \omega}^2 =-1$.
The linearized motion in the vicinity of a phase space
trajectory $\pSpace(t)=[q(t),p(t)]$ is given 
by the \jacobianM\ 
\[
\delta\pSpace(t) = {\bf J}^t(\pSpace)\, \delta\pSpace(0)  
\enspace , \qquad J^t(\xInit)_{ij}
  = {\pde x_i(t) \over \pde \xInit_j}
	\enspace , \qquad \xInit=x(0)
\enspace .
\]
The equations of motion of $\bf J$ follow from \refeq{symplect}
\begin{equation}\label{Bew_Mono}
 {\D \over \D t}{\bf J}^t(\pSpace) 
 = {\bf L}(\pSpace,t) \, {\bf J}^t(\pSpace)
 \enspace , \qquad
\mbox{with }\quad \left. {\bf L}(\pSpace,t)_{mn} = {\bf \omega}_{mk}
               H_{kn} (\pSpace)\right|_{\pSpace(t)} \enspace .
\end{equation}
where $H_{kn}=\partial_k \partial_n H$ is the matrix of second
derivatives of the Hamiltonian.
 $\bf L$ is infinitesimal generator of symplectic 
(or canonical) transformations which leaves ${\bf \omega}$ invariant
\begin{equation} 
 {\bf L}^T {\bf \omega} + {\bf \omega L} = 0
\enspace .
\label{Fun_L}
\eeq

${\bf J}$ is a symplectic matrix, 
as it preserves the symplectic bilinear invariant
${\bf \omega}$:
\begin{equation} 
 \label{M_symb} {\bf J}^T {\bf \omega J} = {\bf \omega}\enspace . 
\end{equation}
{}From this follows that $\det\, {\bf J} =1$, and that the transpose
${\bf J}^T$ and the inverse ${\bf J}^{-1}$ are also symplectic; ${\bf
J}^{-1} = - {\bf \omega}{\bf J}^T{\bf \omega}$.  Hence if $\Lambda$ is
an eigenvalue of $\bf J$, so are $1/\Lambda$, $\Lambda^*$ and
$1/\Lambda^*$.

Let $\jMConfig$ be the {\em configuration} space \jacobianM
\beq
 \jMConfig_{ij}^t(\pSpace) :={\D  q_i (t) \over \D q_j(0)}
 \enspace ,\quad\quad
 \jConfig^t(\pSpace) :=  \det\, \jMConfig^t(\pSpace) 
\enspace ,		
\ee{(6_jac)}
and $\jConfig$ the {\em configuration} space \jacobian\ evaluated on
the $q$-space projection of the phase-space trajectory $\pSpace(t)$
passing through the $t=0$ initial point $\pSpace=(q,p)$.  The
curvature matrix \refeq{curv_matr} is related to the configuration
space \jacobianM\ \refeq{(6_jac)} by
\[
 M_{ij}(\pSpace,t) =  {\partial v_i \over \partial q_j}
  =  {\partial q_k(0) \over \partial q_j(t)}
    {\D~\over \D t}{\partial q_i(t) \over \partial q_k(0)}
  = 
	\left({1 \over \jMConfig^t}\right)_{kj}
	\left({\D~ \over \D t} \jMConfig^t \right)_{ik} \enspace ,
\]
so the configuration space \jacobianM\ satisfies
\beq
 {d~ \over dt} \jMConfig^t = {\bf M} \, \jMConfig^t 
\ee{M_dot}
and is given by the exponentiated time-ordered integral of the trace
of ${\bf M}$
\beq
 \det\, \jMConfig^t(\pSpace)
 =
 \mbox{T} \E^{\int_0^t \D \tau \,\tr{\bf M}^\tau}\enspace .
\ee{lille_jacob}

The {\em full phase space} \jacobianM\ $\jMps$ is given by
\beq
 \VectorII{\delta {\bf q}'}{\delta {\bf p}'} =
 \jMps \VectorII{\delta {\bf q}}{\delta {\bf p}} 
   = \MatrixII{ {\bf J}_{qq} }{ {\bf J}_{qp}}
                     { {\bf J}_{pq} }{ {\bf J}_{pp}}
    \VectorII{\delta {\bf q}}{\delta {\bf p}} 
\enspace ,
\ee{(glob_ja)}
where $\delta {\bf q}$, $\delta {\bf p}$ are $d$-dimensional
infinitesimal tangent space vectors, and ${\bf J}_{qq}$, ${\bf
J}_{qp}$, ${\bf J}_{pq}$ and ${\bf J}_{pp}$ are the $[d\times d]$
submatrices of the full $[2d\times 2d]$ \jacobianM.  (To save paper,
we suppress the $t$, $q$, $p$ dependence for the time being).  Take a
derivative ${\partial / \partial \delta q_i}$ of both sides of
(\ref{(glob_ja)}), keeping terms to linear order in $\delta q$.  This
expresses the configuration \jacobianM\ $\jMConfig$ and the curvature
matrix \refeq{curv_matr} ${\bf M}'$ in terms of the $\jMps $ and the
initial ${\bf M}$
\beq
 \VectorII{\jMConfig}{{\bf M}' \,\jMConfig} =
		\jMps \VectorII{{\bf I}}{{\bf M}}
 \enspace .
 \label{glob_M}
\eeq
Using (\ref{M_dot}) we see that $ \jMps$ evolves the
configuration \jacobianM\ and its time derivative
\[
\VectorII{\jMConfig^t}{{\D ~\over \D t} \jMConfig^t} =
                \jMps \VectorII{{\bf \jMConfig}^0}
       {{\D~\over \D t} \jMConfig^0}
\enspace ,
\] 
where the initial condition for $\jMConfig^0 = {\bf 1}$ for $t=0$.

To spell it out: for a given initial set of $\delta {\bf q}$'s
and $\delta {\bf p}$'s,  the projection of the phase space
volume onto the configuration space 
is given by the configuration space \jacobianM\
$\jMConfig$
\beq
 \jMConfig = \jMConfig^t(q,p,{\bf M})
          :=  {\bf J}_{qq} + {\bf J}_{qp} {\bf M}
 \enspace ,
\ee{15.16a}
and the matrix of curvatures ${\bf M}'$ is evolved recursively by
\beq
{\bf M}'= {\bf M}^t(q_0,p_0,{\bf M}_0)
          :=  \left({\bf J}_{pq} + {\bf J}_{pp} {\bf M}\right)
           {1\over {\bf J}_{qq} + {\bf J}_{qp} {\bf M}} \enspace ,
\ee{15.16b}
where the $q$, $p$, $t$ dependence is hidden in $\jMps$.
We also note that transposing \refeq{glob_M}, multiplying from the right by 
$\omega {\bf J}$, and using the symplectic invariance \refeq{M_symb}
yields
an alternative formula for the configuration space \jacobianM\
\beq
 \left({1\over \jMConfig}\right)^T =
		{\bf J}_{pp} - {\bf M}'{\bf J}_{qp}
\enspace .
\ee{inv_j}

Evaluation of the trace \refeq{Tr_La2} requires a first variation
in all of the dynamical space coordinates $X$, including
$\delta {\bf M}'$. {}From  (\ref{15.16b}) together with 
\refeq{inv_j} we obtain
\beq
 \delta {\bf M}' = 
      {\bf J}_{pp}\, \delta {\bf M}\, {1\over \jMConfig}
                  - {\bf M}'\, {\bf J}_{qp} \,
               \delta {\bf M} {1\over \jMConfig}
	= \left({1\over \jMConfig}\right)^T
		    \delta {\bf M} \,{1\over \jMConfig}
\enspace ,
\ee{var_M}
so the trace \refeq{Tr_La2} is simply reinstated
\beq
\Delta_{p,r} = \sum 
{(\det \, \jMConfig_p)^{r/2}
 \over
  \left|\det\left({\bf 1}
    -{ \pde ~ \over \pde {\bf M} }
           {\bf M}^{\period{p}r} (M) \right)\right|
 }
	= \sum 
{ (\det \, \jMConfig_p)^{r/2}
 \over
  \left|\det\left({\bf 1}
    -\jMConfig_p^{-r} \otimes \jMConfig_p^{-r} \right) \right|
 }
 \enspace .
\ee{trac_M}
The sum is over all ${\bf M}$ that satisfy the fixed point
condition 
\beq
 {\bf M}^\period{p}(q,p,M)={\bf M}
 \enspace .
\ee{fix_M}
Consider now $\jMConfig$ for a periodic orbit $p$; $\jMConfig$ is a
[$d\times d$] matrix with eigenvalues and eigenvectors
\[
 \jMConfig \, {\bf e}_i = \Lambda_i {\bf e}_i
 \enspace ,\qquad	i=1,2,\cdots,d
 \enspace .
\]
Multiply \refeq{glob_M} from the right by the $2d$-dimensional
vector $[{\bf e}_i,{\bf e}_i]$; we see that an eigenvalue of
$\jMConfig$ is also an eigenvalue of the [$2d\times 2d$] phase space
\jacobianM:
\[
 \Lambda_i \VectorII{{\bf e}_i}{{\bf M} {\bf e}_i} =
                \jMps \VectorII{{\bf e}_i}{{\bf M} {\bf e}_i}
 \enspace ,
\]
Furthermore, transposing this equation, multiplying it from
right by $\Lambda_i^{-1}\omega \jMps$, and using 
the symplectic condition $\refeq{M_symb}$ yields the associated
left eigenvector with eigenvalue $1/\Lambda_i$,
\[
 [{\bf e}_i^T, {\bf e}_i^T {\bf M}] \, \omega \Lambda_i^{-1} =
	[{\bf e}_i^T, {\bf e}_i^T {\bf M}] \, \omega \jMps
 \enspace .
\]
In this way the $(\Lambda_i, 1/\Lambda_i)$ pairs of eigenvalues of the
[$2d\times 2d$]-dimensional phase space \jacobianM\ correspond the $d$
eigenvalues of the $d$-dimensional $\jMConfig$.  As the $d$
eigenvalues of $\jMConfig$ generate the $d$ pairs of eigenvalues of
$\jMps$, the sum \refeq{trac_M} gets $2^d$ contributions
$\Lambda_1^{\pm 1} \Lambda_2^{\pm 1} \cdots \Lambda_d^{\pm 1}$.  Each
of these is expanding on its own ${\bf M}$ subspace, and the dominant
one is the most expanding one, so we keep from
\refeq{trac_M} only the modulus of the leading term (the phase will be
treated in the next section) 
\beq
 \left |\widetilde{\Delta}_{p,r} \right |= 
  \prod_{i=1}^d
 { |\Lambda_{p,i} |^{r/2}  
 \over
 1 -{ 1 / \Lambda_{p,i}^{2r} }
 }
 \enspace .
\ee{trac_M_j}
The dynamics in the tangent space can be restricted to a 
unit eigenvector neighborhood corresponding to the largest eigenvalue
of the \jacobianM. On this neighborhood the largest eigenvalue of the
\jacobianM\ is the only fixed point, and the \Vd\ obtained by keeping
only the largest term in the $\Delta_{p,r}$ sum in \refeq{trac_M} is
also entire, \cite{CV93}.

So, (very pleasantly) as $\Lambda_i$ are also eigenvalues of the
configuration space \jacobianM\ $\jMConfig$, the extra trace over
${\bf M}$ is coming for free; we have {\em already} computed the
eigenvalue set $\{\Lambda_1, 1/\Lambda_1, \cdots, \Lambda_d,
1/\Lambda_d\}$ for every full $(q,p)$ phase space cycle $p$.

\section{Maslov Indices}
\label{s_Maslov}

The square root of the configuration space \jacobian\
\refeq{lille_jacob} is also a time ordered integral
\begin{equation}
 \left(\det\,\jMConfig^t(x)\right)^{\half}  =
 \mbox{T} \exp\left\{\frac{1}{2}
 \int_{0}^{t}\D\tau \,\tr\left(
 {\bf M}^\tau\right)\right\} \enspace .
 \label{xx}
\end{equation}
{\bf M} diverges at {caustics}; for example, for $d=1$ Poincar\'{e}
sections (such as for billiards) ${\bf M}= {\pde p / \pde q}$ diverges
whenever a trajectory points in the $p$-axis direction.  Close to a
singularity, where
$$
 {\bf M}(t\rightarrow t^c)=\infty 
  \enspace , 
$$
we can neglect the non-leading terms from \refeq{M_equat} 
and use the solution of
\beq
 \dot{{\bf M}}=-{\bf M}^2
 \enspace ,
 \ee{singsing}
after 
the symmetric matrix ${\bf M}$ had been transformed into a diagonal
form. 
The time ordered integral close to the singularity is dominated by
\[
\left( \det \left\{ {\D q_i (t^c_{+}) \over \D q_j(t^c_{-})}\right\} 
\right)^{1/2}
  =
 \exp\left(\frac{1}{2}\int_{t^c_{-}}^{t^c_{+}}
 \frac{R}{\tau+\I \epsilon -t_c}\D\tau\right)
 \enspace ,
\]
where $t^c_{\pm} = t^c \pm \eta$ are infinitesimally close to
$t^c$ and the integration variable $\tau$ is shifted to $\tau+\I\epsilon$,
because the corresponding wave packet should start out with a positive phase
before it encounters the first singularity. 
This integral can be computed by taking the limit $\epsilon\rightarrow 0$, 
\beq
\left( \det \left\{ {\D q_i (t^c_{+}) \over \D q_j(t^c_{-})}\right\} 
\right)^{1/2}=\exp(-\I\pi(R/2)) 
\left |  \det \left\{ {\D q_i (t^c_{+}) \over \D q_j(t^c_{-})}\right\} 
 \right |^{\half}
\enspace .
\ee{Maslov_ph}
Note that the phase only results from the delta function part 
of the integrand, whereas the principle value 
contributes  just to the modulus which
has been already calculated in \refeq{trac_M_j}.  
Between two singular points the time ordered integral is positive and
gives the absolute value of the volume ratio.  $R$ counts the number
of rank reductions of the matrix ${\bf M}$ along the classical path,
and it is a function of the initial condition ${\bf M}_0$; for a
periodic orbit it is an invariant property of the cycle.

\section{Gutzwiller Trace Formula vs. Quasiclassics}
\label{s_GV_vs_quasi}

Consider a generalization of the
\Vd\ \refeq{Quasi-det}, weighted by extra powers of $\Lambda_{p,i}$:
\beq
 F_{n}(k) =\exp\left(-\sum_{p}\sum_{r=1}^\infty
        {1 \over r}
	\prod_{i=1}^d
      {  { |\Lambda_{p,i}|^{-r/2}\,   \E^{\frac{\I}{\hbar}  S_p(k) r
         -  \I \pi \frac{\maslov{p}}{2}r}}
	\over
         \Lambda_{p,i}^{n r} 
     (1-1/\Lambda_{p,i}^r)^2 (1-1/\Lambda_{p,i}^{2r})}
 \right) \enspace .
\label{Vatt-det}
\eeq
The weight $1/(1-x)$, $x=1/\Lambda_{p,i}^r$ 
of $p$-th term in the exponent of the \qS\ \refeq{GV_zeta}
can be related to the \Vd\ cycle weight $1/(1-x)^2 (1-x^2)$
in  \refeq{Vatt-det} by multiplying it by
\[
 1= {1\over (1-x)(1-x^2)} 
  - {x\over (1-x)(1-x^2)} - {x^2\over (1-x)(1-x^2)}
   + {x^3\over (1-x)(1-x^2)}
\enspace  .
\]
{}From this it follows that the \qS\ function \refeq{GV_zeta} for Axiom
A flows is meromorphic in the complex $k$ plane, as it can be written
as a ratio of entire functions; for 2-dimensional Hamiltonian systems
\beq
\Zqm(k) = { F_{0}(k) F_{3}(k)
              \over 
              F_{1}(k) F_{2}(k)  }
 \enspace ,
\ee{Z-GV}
where $F_{k}(k)$ includes only \refeq{trac_M_j},
the first term in the $\Delta_{p,r}$
sum \refeq{trac_M}.  The zeros of the \qS\ coincide with the ones
obtained from $F_{0}(k)=\Zqc(k)$, and the leading poles should arise from
$F_{1}(k)$.
In two dimensions, i.e.\ $d=1$, \refeq{Vatt-det} can be resummed as 
\beq
F_{n}(k)
  = \prod_p \,\prod_{j=0}^{\infty}\, \prod_{l=0}^{\infty}\,
  \left(1 -\frac{t_p}{\Lambda_p^{n+j+2l}}\right )^{j+1}
  \enspace ,
  \label{Vatt-det-twodim}
\eeq
where $t_p$ is defined in \refeq{t_p_def}.

\section{Selberg Zeta Function}
\label{App_Selberg}

The question that arises naturally in discussing semiclassical
quantization is following: if the usual semiclassical evolution is not
multiplicative, why does it anyway yield the {\em exact} quantization
in the case of the Selberg trace formula? And what does the
quasiclassical quantization yield for flows on surfaces of constant
negative curvature?

The Selberg  (1956) zeta function		
for geodesic flows on surfaces of constant negative curvature is
exceptional: in this very special case the multiplicativity is
guaranteed by the Bowen-Series (1979) map, which reduces the
two-dimensional flow to a direct product of 1-dimensional maps, and
makes it possible to construct the associated transfer operators in
terms of one variable, \cite{Mayer90:93}.

The essence of the construction is the following:
In the Poincar\'e halfplane representation the dynamics
is described by the free Hamiltonian
\begin{equation}
 \hat{H}=\frac{1}{2y^2}(p_x^2 + p_y^2)
\end{equation}
whose classical trajectories are circle segments.  The centers of the
circles always lie on the $y=0$ axis, and any free trajectory can be
characterized by $x_f$ and $x_b$, the forward and backward
intersection points of its circle with the $y=0$ axis.  The polygonal
billiards in the $x,y$ plane are defined in terms of walls which
themselves are geodesics, hence also characterized by their footpoints
$x_1,x_1'$, $x_2,x_2'$ ... (see \reffig{f_selberg}).
%
\begin{figure}
\centerline{\epsfig{file=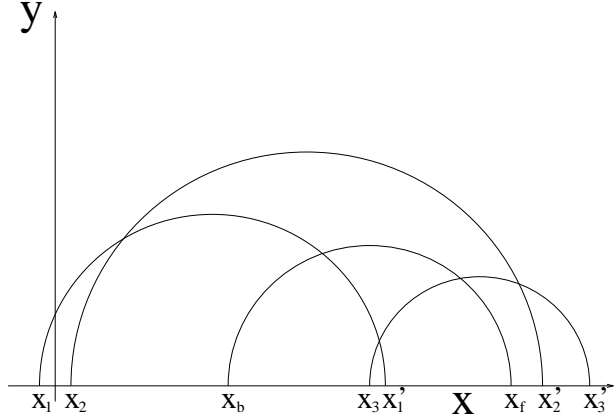,width=8cm,angle=-90}}
\caption[]{A typical arrangement on the Poincar\'e halfplane.
The half circles with footpoints $(x_i,x_i')$, $i=1,2,3$ are the 
billiard walls. The forward and backward footpoints $(x_b,x_f)$
represent a trajectory.}
\label{f_selberg}
\end{figure}
%
A reflection off a wall changes the direction of the particle, with
the new trajectory characterized by a new pair of footpoints $x_f'$
and $x_b'$.  The new forward footpoint will be the image of the old
footpoint with respect to an inversion transformation on the circle of
the wall.  For example, a reflection off the wall $x_n,x_n'$ of radius
$R_n=|x_n-x_n'|/2$ and center $x_n^c=(x_n+x_n')/2$ is described by
\[
 x_f'=f_n(x_f)=x_n^c+R_n^2/(x_f-x_n^c)
\enspace .
\]
The forward footpoint and the index of the wall determine uniquely the
next forward footpoint. The footpoint of a periodic orbit reflected
off walls $\epsilon_1\epsilon_2...\epsilon_{n_p}$ respectively is
determined by the equation
\begin{equation}
 x_p=F_{\epsilon_1\epsilon_2...\epsilon_{n_p}}(x_p)=
 f_{\epsilon_{n_p}}(f_{\epsilon_{n_p -1}}
 (...f_{\epsilon_1}(x_p) ...))\enspace .
\end{equation}
The hyperbolic length of this periodic orbit is
$l_{\epsilon_1\epsilon_2...\epsilon_{n_p}}=
\log |F'_{\epsilon_1\epsilon_2...\epsilon_{n_p}}(x_p)|$,
and its stability eigenvalue is also given by the derivative
$F'_{\epsilon_1\epsilon_2...\epsilon_{n_p}}(x_p)$.  
The stability is
the product of derivatives evaluated along the orbit
\[
   F'_{\epsilon_1\epsilon_2...\epsilon_{n_p}}
   = 
  \prod_{i=1}^{n_p} F'_{\epsilon_i}
\enspace ,
\]
and is multiplicative without any need for further manipulations.
This property makes the polygonal billiards on surfaces of constant
negative curvature unique and atypical.

The Fredholm determinant of the 1-dimensional Perron-Frobenius operator
\[
 {\cal L}(y,x,k)=|f'(x)|^{1/2 +\I k}\delta(y-f(x))\enspace ,
\]
where $f$ is the appropriate footpoint mapping and $k=\sqrt{E-1/4}$ is
the wave number, is precisely the Gutzwiller-Voros \qS\ for this
problem, $\Zqm(E)=\det(1-{\cal L}_0)$.  Unlike the generic situation
discussed in this paper, the \qS\ is in this case an entire function.
However, the spectrum of the \Vd\ $\Zqc(E)$ defined in this paper
contains spurious zeroes in the complex plane in addition to the true
zeroes on the real $k$ axis. These spurious zeroes are the eigenvalues
of weighted operators of type
\begin{equation}
 {\cal L}_m(y,x,k)=|f'(x)|^{1/2+m +\I k}\delta(y-f(x))\enspace ,
\end{equation}
where $m$ is an integer number. Since the $F_{n}(k)$'s (see 
\refeq{Vatt-det-twodim}) 
can be expressed in
terms of the \Fd s of these operators as
\beq
 F_{n}(k) = \prod_{l=0}^{\infty} \, \bigg \{ 
 \det\left(1- {\cal L}_{n+l}\right) \,\det\left(1- {\cal L}_{n+l+1} \right )
      \bigg \}^{l+1}
\enspace ,
\ee{Selb_Zqc}
$\Zqm(k)=\det(1-{\cal L}_0)$ results under the relation
\refeq{Z-GV}, too.

\end{document}